\documentclass[final]{siamltex}
\usepackage{lineno}
\usepackage{algorithmic} 
\usepackage{algorithm}

\usepackage{float}
\usepackage{paralist}
\usepackage{lipsum}
\usepackage{subfigure}

\usepackage{amssymb}
\usepackage{wrapfig}
\usepackage{graphicx}
\usepackage{color}
\usepackage{easylist}
\usepackage{amsmath}
\usepackage{lineno}

\usepackage{easylist}

\renewcommand{\vec}[1]{\ensuremath\boldsymbol{#1}}

\newcommand{\gphi}{\nabla\phi}

\title{ Electrohydrodynamics of Three-Dimensional Vesicles: A Numerical Approach \footnotemark[3]}

\author{Ebrahim M. Kolahdouz\footnotemark[1] \and David Salac\footnotemark[1]\ \footnotemark[2]}

\begin{document}

	\maketitle

	\renewcommand{\thefootnote}{\fnsymbol{footnote}}
	\footnotetext[1]{University at Buffalo, Department of Mechanical and Aerospace Engineering,	Buffalo, NY, 14260}
	\footnotetext[2]{Corresponding Author {\tt davidsal@buffalo.edu}}
	\footnotetext[3]{This work supported by NSF Grant \#1253739}
	\renewcommand{\thefootnote}{\arabic{footnote}}

	\begin{abstract}
	
		A three-dimensional numerical model of vesicle electrohydrodynamics in the presence of DC electric fields
		is presented. The vesicle membrane is modeled as a thin capacitive interface through the use
		of a semi-implicit, gradient-augmented level set Jet scheme. The enclosed volume and surface area are conserved both locally
		and globally by a new Navier-Stokes projection method. The electric field calculations 
		explicitly take into account the capacitive interface by an implicit Immersed Interface Method
		formulation, which calculates the electric potential field and the trans-membrane potential 
		simultaneously. The results match well with previously published experimental, analytic
		and two-dimensional computational works. 
		
	\end{abstract}

	\begin{keywords}
		Vesicle, Stokes flow, Level set, Continuous surface force, Numerical method
	\end{keywords}

%%%%%%%%%%%%%%%%%%%%%%%%%%%%%%%%%%%%%%%%%%%%%%%%%%%%%%%%%%%%%%%%%%%%%%%%%%%%%%%%%%%%%%%%%%%%%%%%%%%%%%%%%%%%%%%%%%%%%%%%%%%
\section{Introduction}
\label{sec:1.0}

Giant liposome vesicles are enclosed bag-like membranes composed of lipid bilayers. 
These vesicles share many similarities in composition and size with other biological cells such as red blood cells. 
Experimental observations have demonstrated that lipid vesicles exhibit a striking resemblance to more complicated biological cells in terms of their 
equilibrium shapes and dynamic behavior in various fluid conditions \cite{Abkarian2008,Vlahovska2009}.
These make liposome vesicles a robust model system to study the behavior of more 
complicated non-nucleated biological cells.

It is also quite easy to artificially create these soft particles in a laboratory setting; in an aqueous solution
lipid molecules will self-assemble into two-dimensional sheets, which then 
fold and curve in three-dimensional space and form a fluid-filled
vesicle \cite{lasic1988mechanism}. The ease of liposome vesicle formation, in addition to their bio-compatibility, has led to vesicles 
being proposed as building blocks for various biotechnologies such as directed drug and gene 
delivery \cite{Allen2004,Jesorka2008} or biological microreaction \cite{Noireaux2004,Tanner2011}.

Vesicles interacting with external flows have been of major recent interest.
Three well-known types of vesicle dynamics in shear flow are tank-treading, tumbling and a transient state called trembling or vacillating breathing.
These behaviors have been extensively studied in theory \cite{helfrich1973,seifert1991, miao1994,lebedev2007,vlahovska2007} and experiments \cite{abkarian2002,kantsler2006,Haas1997}.
Several numerical simulations of vesicle dynamics have also appeared in the literature.
Among the numerical studies in two dimensions are models using the boundary integral method \cite {Beaucourt2004,veerapaneni2009}, the phase field \cite{biben2003, maitre2012, Aland2013}, 
a coupled level set and projection method
\cite{salac2011, salac2012}, a coupled level set and finite-element method \cite{Doyeux2013251,laadhari2012} and the lattice Boltzmann method \cite{Kaoui2011}.
A few three-dimensional studies have been also reported using the phase-field approach \cite{biben2005}, 
boundary integral method \cite{Biben2011, ZHAO2011,Veerapaneni2011} and front-tracking \cite{Yazdani2012}.
Studies have shown the dynamics of the vesicle depends on three major parameters: the viscosity ratio between the enclosed and surrounding fluids, 
the shear rate and the reduced-volume, which is defined as the ratio of the volume of the vesicle to the volume of a sphere with the same surface area as the vesicle.

The use of electric fields, in combination with fluid flow, has been proposed as a powerful method to direct the behavior of vesicles
towards a wide range of biotechnological applications. 
Weak electric fields have found applications in cell manipulation techniques 
such as electrofusion \cite{wang2013}, tissue ablation \cite{campus2006}, wound healing \cite{keese2004}, and in the treatment of tumors \cite{skog2008}.
Strong electric fields induce electro-poration in vesicles through the formation of transient pores in the membrane and could play a role 
in novel biotechnological advances such as the delivery of drugs and DNA into living cells \cite{neumann1982,tsong1991}.

Recent experiments have reported on the topological behavior of vesicles subjected
to either a DC or AC electric field with different intensity, frequency
and duration of exposure to the field \cite{Riske2005,Riske2006,dimova2007,salipante2014}.
Experiments show that depending on the conductivity and permittivity differences between the enclosed and surrounding fluids the vesicle undergoes various 
shape transformations, such as prolate (major-axis aligned with the electric field)
to oblate (major axes are perpendicular to the field).
One very interesting phenomena in this context is the dynamics of an initial prolate vesicle in a strong DC field with a small enclosed fluid conductivity as
compared to the conductivity of the surrounding fluid.
It is theoretically expected that as time progresses
the vesicle transitions to an oblate shape first, then evolves back
into the prolate shape. During such transition 
nearly cylindrical shapes with high-curvature edges were observed for a vesicle subjected to strong pulses \cite{Riske2006}. This behavior was attributed to the 
presence of salt in the solution and not due to the membrane characteristics. However in \cite{Salipante2013thesis} the poration of the vesicle membrane was 
proposed as the possible explanation for both vesicle collapse and cylindrical deformations.

In addition to recent experimental works, theoretical models have also investigated the 
electrohydrodynamics of nearly-spherical vesicles. Leading-order perturbation analysis has been employed to
obtain reduced models in the form of ordinary differential equations \cite{vlahovska2009elec, schwalbe2011,schwalbe2011lipid,seiwert2012stability}.
In another work a spheroidal shell model has been used to investigate the morphological change of the vesicle in AC electric field \cite{Yugang2013}.

Despite numerous theoretical investigations, numerical studies of the vesicle electrohydrodynamics are rare. In a recent work, a boundary integral 
method was employed to study different equilibrium 
states of a two-dimensional vesicle in the presence of a uniform DC electric field \cite{mcconnell2013}. 
A Prolate-Oblate-Prolate (POP) transition
was captured for a vesicle with an inner-fluid conductivity smaller than the surrounding region.

Currently, there is still a gap between the morphological changes observed in experiments, what theoretical models predict, and 
methods to control this behavior. This gap needs to be addressed before vesicles can form a building block for electrohydrodynamic based microfluidic systems and technologies.
Part of the difficulty arises from modeling the complex physics of the vesicle electrohydrodynamics and challenging phenomena such as membrane poration or fusion.
Compared to hydrodynamics investigations, the electrohydrodynamics of the vesicle has substantially faster dynamics with much larger deformations which makes the numerical modeling nontrivial.
Hence a lack of thorough numerical investigation with different material properties and electric field parameters still remains.

In a previous work by the authors the Immersed Interface Method (IIM) was utilized to solve for both the electric potential and trans-membrane potential 
around a vesicle in three dimensions \cite{Kolahdouz2013}. The jump conditions for
potential and its first and second derivatives on the interface were determined and utilized to obtain accurate
electric potential solutions for a three-dimensional vesicle with an arbitrary shape.
In this paper the electric field model presented in this recent work is combined with a projection-based hydrodynamics solver and a 
semi-implicit jet scheme for capturing the interface.
This model is used to study the electrohydrodynamics of three dimensional vesicles in general flow.  
The dynamics of the vesicle is determined by the interplay between the hydrodynamics, bending, tension and electric field stresses on the
membrane. This is one of the first attempts at investigating the 
electrohydrodynamics of vesicles in three-dimensions. As one may expect, for this kind of 
physics a three dimensional model will result in a much richer and wider range of topological changes.
The differences compared to a two dimensional model will be subtle and important at the same time.  
In addition to this, a full Navier-Stokes system of equations is considered here 
unlike all the previous models which investigate the problem in the Stokes region.
For vesicles in strong DC field, the approximate velocity may sometimes exceed $0.01$ m/s \cite{Salipante2013thesis}, resulting in non-trivial Reynolds numbers, and therefore,
the Stokes assumption for the fluid flow might not always hold.
Moreover, the method presented here allows for studying deflected vesicles 
with smaller reduced volumes than what theoretical models normally address and is able to predict the type of deformations observed in experiments.

The remainder of this paper is organized as follows. In Section \ref{sec:2.0} the physical system and formulation behind the electrohydrodynamics of the vesicle are described.
In Section \ref{sec:3.0} the numerical algorithm is presented. This will be followed by sample numerical results in Section \ref{sec:4.0}. Final
remarks and possible future work then follow.

%~ while vesicles and cell membranes are
%~ essentially incompressible,1 with no
%~ possible transport of molecules between
%~ the membrane and the inner volume,
%~ ensuring a constant membrane area.

%%%%%%%%%%%%%%%%%%%%%%%%%%%%%%%%%%%%%%%%%%%%%%%%%%%%%%%%%%%%%%%%%%%%%%%%%%%%%%%%%%%%%%%%%%%%%%%%%%%%%%%%%%%%%%%%%%%%%%%%%%%
\section{Theory and formulation}
\label{sec:2.0}

Let a three-dimensional vesicle of encapsulated volume $V$ have a surface area of $A$. Deviation from a perfect sphere is measured
by a reduced volume parameter, $v$, defined as the ratio of the vesicle volume to the volume of a sphere with the same surface area: $v = 3V/(4\pi a^3)$ 
where $a = \sqrt{A/4\pi}$ is the characteristic length scale. The typical size of the vesicle is $a \approx 10-20\;\mu$m while the thickness of the 
bilayer membrane is $d \approx 5$ nm \cite{Seifert1997}. Due to the three-orders of magnitude difference 
between the vesicle size and the membrane thickness the membrane will be treated as an infinitesimally thin interface 
separating the inner and outer fluids.
The vesicle membrane is assumed to be impermeable to fluid molecules and the number of
lipids on the membrane does not change over time, while the surface density of lipids at room temperature
is constant \cite{Seifert1997}. These two conditions result in an inextensible membrane with 
constant enclosed volume and global surface area, along with local surface incompressibility.
Another important feature of the membrane is its electrical insulating property and impermeability against ionic transfer \cite{salipante2014,vlahovska2009elec,mcconnell2013}.
Therefore, in the presence of an external electric field
the membrane acts as a capacitor. This capacitive property is an important factor in studying the dynamics of vesicles in the presence of electric field.

\subsection{Electric Field Equations}
\label{sec:2.1}

\begin{figure}[ht!]
	\centering
	\includegraphics[width=\textwidth]{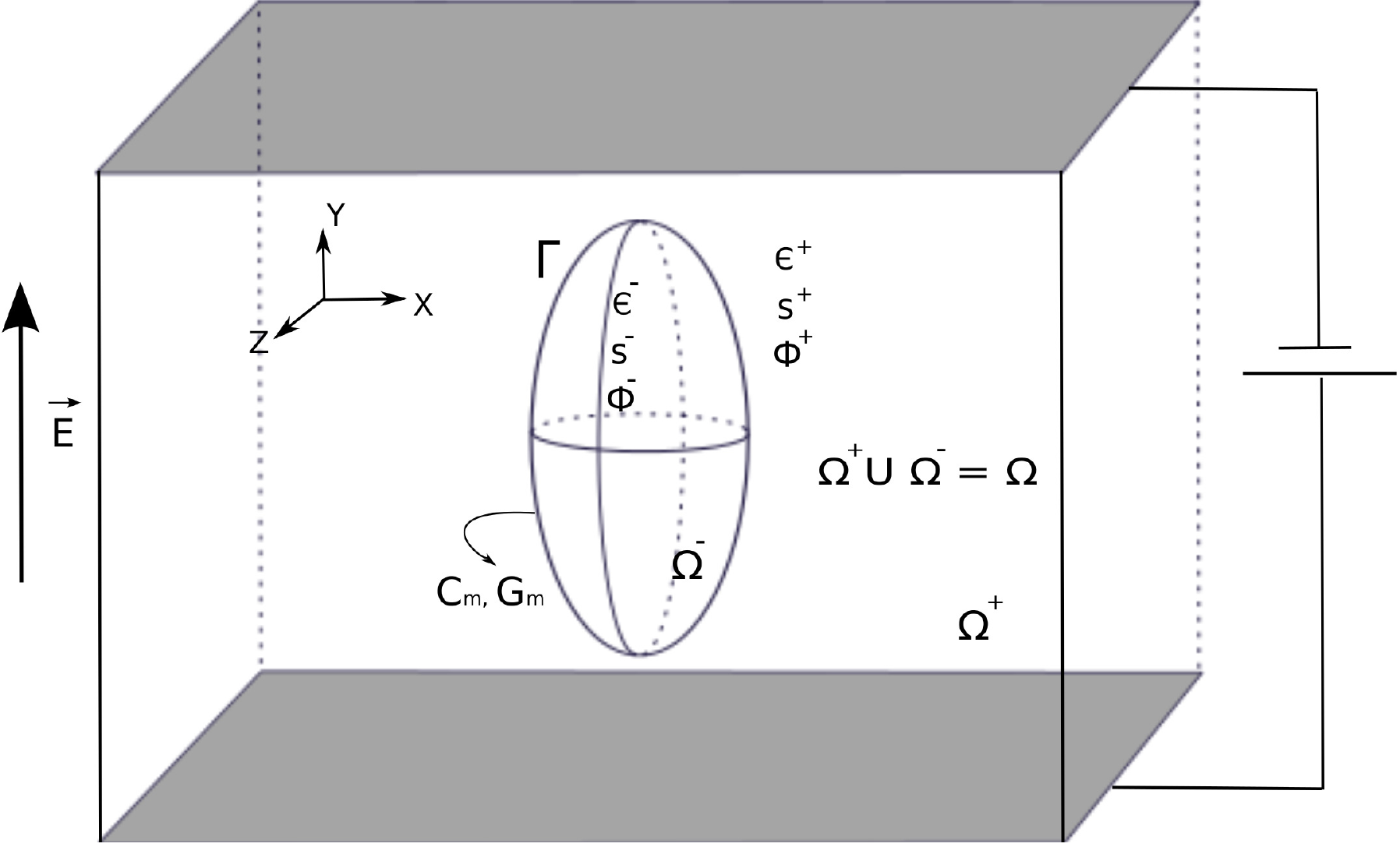}
	\caption{Schematic of a vesicle subjected to external electric field }	
	\label{fig:drawingEHD}
\end{figure}

A schematic of a vesicle exposed to an external electric field is illustrated in Fig. \ref{fig:drawingEHD}. Different 
properties inside $(-)$ and outside $(+)$ of the membrane are shown in the figure. 
The embedded fluid ($\Omega^-$) is separated from the surrounding fluid ($\Omega^+$) by the vesicle 
membrane represented as $\Gamma$. This membrane is assumed to be made of a charge-free lipid bilayer with uniform and constant capacitance $C_m$ and conductivity $G_m$.
The vesicle is suspended in a fluid of conductivity $s^+$ and permittivity $\epsilon^+$. 
The enclosed region is assumed to have a different conductivity $s^-$ and permittivity  $\epsilon^-$.  
It is assumed that properties are constant in each fluid, with a finite jump occurring across the membrane.
%~ Within each domain, there exist velocity ($\vec{u}$), pressure ($p$), viscosity ($\mu$), permittivity($\epsilon$) and electric conductivity ($s$).
%~ The membrane i

The leaky-dielectric model is assumed to hold in the whole domain \cite{Saville1997}.
This results in no local free charge in the bulk fluids, and thus the electric field ($\vec{E}$) is irrotational and given 
as the negative gradient of the electric potential ($\Phi$)
\begin{equation}
	\vec{E}^k=-\nabla\Phi^k,
	\label{eq:Elec_field} 
\end{equation}
where $k=-$ or $+$ denotes the fluid domain.
The lack of free charge implies that the electric potential is a solution of Laplace equation in each domain
\begin{align}
	\nabla^2\Phi^k&=0.
	\label{eq:Elec_lap}   
\end{align}
However, the presence of the capacitive interface leads to a discontinuity in the potential across the membrane
\begin{equation}
	\Phi^{+}-\Phi^{-}=-V_m,
	\label{eq:Phi_Difference}
\end{equation}
where $V_m$ is the transmembrane potential and can be obtained from the conservation of current density across the membrane \cite{seiwert2012stability,DeBruin1999},
\begin{equation}
	C_m\frac{dV_m}{dt}+G_{m}V_{m}=\vec{n}\cdot\left(s^{k}\vec{E^{k}}\right)+\epsilon^\pm\dfrac{\partial \vec{E}^\pm}{\partial t}\cdot\vec{n}+\nabla_s\cdot\left(\vec{u}Q^\pm\right),	
	\label{eq:Vm_equation_simplified} 
\end{equation}
where $\vec{n}$ is the outward facing normal vector on the interface (pointing into the $\Omega^+$ fluid), $\nabla_s$ is the surface gradient operator, $\vec{u}$ 
is the underlying fluid velocity and $Q^\pm$ is the induced charge on the top and bottom side of the membrane. In the limit
of fast bulk charge relaxation and negligible charge convection the time evolution of $\vec{E}^\pm$ and the surface convection of $Q^\pm$ can be ignored \cite{seiwert2012stability}.
Furthermore, 
continuity of the Ohmic current $\vec{J}=s\vec{E}$ in the normal direction across the membrane can be written as
\begin{equation}
	\vec{n}\cdot(\vec{J^{+}} - \vec{J^{-}})=\vec{n}\cdot(s^{+}\vec{E^{+}} - s^{-}\vec{E^{-}})=0.
	\label{eq:current_denisity_continuity}
\end{equation}

Due to the difference in electrical conductivities of the inner and outer fluids charges accumulate on the interface at different rates.
The effective charge density, $Q$, induced on the membrane as a result of an external electric field is then expressed as the jump in the normal 
component of displacement field $\vec{D}=\epsilon\vec{E}$:
 \begin{equation}
	\vec{n}\cdot(\epsilon^{+}\vec{E^{+}} - \epsilon^{-}\vec{E^{-}})= Q.
	\label{eq:discontinuity_Q}
\end{equation}

The electric force $\vec{\tau}_{el}$ acting on the membrane is obtained from the Maxwell Tensor $\vec{T}_{el}$:
\begin{align}
	 \vec{\tau}_{el}=\vec{n}\cdot[\vec{T}_{el}], \hspace{3mm} 
	\vec{T}_{el}=\epsilon(\vec{E}\vec{E} -\frac{1}{2}\left(\vec{E}\cdot\vec{E}\right)\vec{I}) \hspace{3mm} \textnormal{on} \hspace{3mm}  \Gamma,
	\label{eq:Maxwell}
\end{align}
where $[\;]$ in the above relation and through the rest of this paper shows the jump across the interface.
For an arbitrary quantity $f$ at location $\vec{x}_{\Gamma}$ on the interface a jump is mathematically defined as
\begin{equation}
	[f] = f^+ - f^- =\lim_{\delta \to 0^+} f(\vec{x}_{\Gamma} + \delta \vec{n}) - \lim_{\delta \to 0^+} f(\vec{x}_{\Gamma} - \delta \vec{n}).
	\label{eq:jump_definition}
\end{equation}
Note that when discussing jumps in quantities at the interface a superscript $+$ or $-$ does not denote a bulk value, but the value as one approaches the 
interface.

%~ As we will see in section (\ref{sec:3.1}) it is worthwhile for the computational porpuse  to rewrite the electric field force in terms of tangential and normal components of the electric field. 
%~ Let us introduce the vectors $\vec{t}$ and $\vec{b}$ as the unit tangent vectors along with unit normal vector $\vec{n}$ on an interface point. Then the electric stress can be rewritten as
%~ 
%~ \begin{equation}
	%~ \vec{\tau}_{el} = (\vec{\tau}_{el}\cdot\vec{n})\vec{n} + (\vec{\tau}_{el}\cdot\vec{b})\vec{b} + (\vec{\tau}_{el}\cdot\vec{t})\vec{t}
	%~ \label{eq:electric_stress}
%~ \end{equation}
%~ 
%~ These three components are given as \cite{yamamoto2010}
%~ 
%~ \begin{align}
	%~ \vec{\tau}_{el}\cdot\vec{n} = \frac{\epsilon}{2}[(\vec{E}\cdot\vec{n})^2 - (\vec{E}\cdot\vec{b})^2 -(\vec{E}\cdot\vec{t})^2]\\
	%~ \vec{\tau}_{el}\cdot\vec{b}= [\epsilon\vec{E}\cdot\vec{n}]\vec{E}\cdot\vec{t}\\
	%~ \vec{\tau}_{el}\cdot\vec{t}= [\epsilon\vec{E}\cdot\vec{n}]\vec{E}\cdot\vec{b}
	%~ \label{eq:electric_stress_nbt}
%~ \end{align}

%~ There are several different time scales associated with vesicle in electric  fields 

\subsection{Fluid Flow Equations}
\label{sec:2.2}
Assume that both bulk fluids are Newtonian and incompressible with a matched density $\rho$. Consequently they both satisfy the Navier-Stokes equations
\begin{align}
	\rho\frac{D\vec{u}^{\pm}}{Dt} = \nabla\cdot\vec{T}_{hd}^{\pm} \hspace{3mm} \textnormal{with} \hspace{3mm}
	\nabla\cdot\vec{u}^{\pm}=0 \hspace{3mm} \textnormal{in} \hspace{3mm}  \Omega^{\pm},
	\label{eq:stokes_compact}
\end{align}
where $\vec{u}$ is the velocity vector, $D/Dt$ is the material (total) derivative, and  $\vec{T}_{hd}$ is the bulk hydrodynamic stress tensor defined as
\begin{equation}
	 \vec{T}_{hd}^{\pm}=-p^{\pm}\vec{I}+\mu^{\pm}(\nabla\vec{u}^{\pm}+\nabla^T\vec{u}^{\pm}) \hspace{3mm} \textnormal{in} \hspace{3mm}  \Omega^{\pm}.
	\label{eq:hydrodynamic_stress}
\end{equation}

 %~ \begin{figure}[ht!]
	%~ \centering
%~ 
	%~ \includegraphics[width=0.6\textwidth]{SketchVesicle}
	%~ \caption{Vesicle configurationInside and outside regions of the vesicle with different physical properties }	
	%~ \label{fig:SchematicVesicle}
%~ \end{figure}

The fluid flow in each region is coupled via the conditions on the inextensible membrane. The velocity is assumed
to be continuous on the surface, $[\vec{u}]=0$.
However, due to the forces exerted by the membrane the hydrodynamic stress undergoes a jump across the interface of the two fluids.
This condition is obtained by balancing the hydrodynamic and electric stresses with the bending and in-extension tractions of the membrane,
\begin{equation}
	\vec{\tau}_{hd} + \vec{\tau}_{el} = \vec{\tau}_{m} + \vec{\tau}_{\gamma} \hspace{3mm} \textnormal{on} \hspace{3mm} \Gamma,
	\label{eq:surf_cond}
\end{equation}
where $\vec{\tau}_{hd} =\vec{n}\cdot[\vec{T}_{hd}]$ is the normal component of the hydrodynamic stress while $\vec{\tau}_{m}$ and $\vec{\tau}_{\gamma}$
are the bending and tension traction forces, respectively.
These two forces together constitute the total membrane force per unit area and are calculated by taking 
the variational derivative of the total energy of the membrane \cite{helfrich1973}. 
For a three dimensional vesicle and neglecting spontaneous curvature the ultimate forms of the bending and tension stresses are found to be \cite{seifert1999fluid}
\begin{align}
	\vec{\tau}_{m} &= -\kappa_c(\frac{H^3}{2}-2HK + \nabla^2_{s}H)\vec{n}, \hspace{3mm} \hspace{3mm}
	\vec{\tau}_{\gamma} =\gamma H\vec{n} -\nabla_s\gamma.
	\label{eq:curvatures}
\end{align}
In this relation, $H = \nabla\cdot\vec{n}$ is twice the mean curvature (the total curvature), 
$K=\nabla\cdot[\vec{n}\nabla\cdot\vec{n} + \vec{n}\times(\nabla\times\vec{n})]$ is the Gaussian curvature, 
$\kappa_{c}$ is the bending rigidity, $\kappa_{g}$
is the Gaussian bending rigidity, $\gamma$ is the tension and $\nabla_{s}=\vec{P}\nabla$ where $\vec{P}=(\vec{I}-\vec{n}\otimes\vec{n})$ is the surface projection operator.

%Applying the identity in (\ref{eq:curvatures}) together with the definitions given for hydrodynamics and electric stresses, the jump condition in (\ref{eq:surf_cond}) can be rewritten as
%\begin{equation}
	%\vec{n}\cdot[-p\vec{I}+\mu(\nabla\vec{u}+\nabla^T\vec{u})] +\vec{n}\cdot[\epsilon(\vec{E}\vec{E} -\frac{1}{2}\vec{E}\cdot\vec{E}\vec{I})] = -\kappa_c(\frac{H^3}{2}-2HK + \nabla^2_{s}H)\vec{n} + \gamma H\vec{n} -\nabla_s\gamma.
	%\label{eq:Final_jump}
%\end{equation}

The last condition to consider is the local area incompressibility constraint on the membrane. 
This condition is enforced by ensuring that the velocity is surface-divergence free on the membrane:
\begin{equation}
	\nabla_s\cdot\vec{u}=0 \hspace{3mm} \textnormal{on} \hspace{3mm}  \Gamma.
	\label{eq:surf_div_free}
\end{equation}

Note that the divergence conditions, $\nabla\cdot\vec{u}=0$ and $\nabla_s\cdot\vec{u}=0$, should be sufficient to conserve both volume and area.
However, in practice it is difficult to ensure that global surface area and enclosed volume are sufficiently conserved \cite{Aland2013}. Therefore, when developing
the numerical method in Sec. \ref{sec:3.0} global conservation will be explicitly enforced as well.

%%%%%%%%%%%%%%%%%%%%%%%%%%%%%%%%%%%%%%%%%%%%%%%%%%%%%%%%%%%%%%%%%%%%%%%%%%%%%%%%%%%%%%%%%%%%%%%%%%%%%%%%%%%%%%%%%%%%%%%%%%%
\subsection{Description of the Interface}
\label{sec:2.3}

The motion of the interface is tracked using a gradient-augmented level set method. First developed by Osher and Sethian \cite{Osher1988} the level set method is based upon an
implicit representation of an interface as the zero level set of a higher dimensional function. This was later extended to explicitly 
include information about the gradients of the level set \cite{Nave2010}. The use of gradient information allows for the accurate determination of 
the interface location and curvature information away from grid node locations \cite{Nave2010}.

Using the same notation for the interface as Fig. \ref{fig:drawingEHD} the
level set function $\phi$ is defined as
\begin{equation}
	\Gamma(t)=\left\{\vec{x}:\phi(\vec{x},t)=0\right\},
\end{equation} 
while the level set gradient field is defined as
\begin{equation}
	\vec{\psi}=\nabla\phi.
\end{equation}
The level set value is chosen to be negative in the $\Omega^-$ and positive in the $\Omega^+$ domain. 
This representation has the advantages of treating any topological changes naturally without
complex remehsing, and the ability to calculate geometric quantities from the level set function and its derivatives. 
For example, the unit normal to the surface, $\vec{n}$, can be easily computed as
\begin{equation}
	\vec{n}=\frac{\vec{\psi}}{\|\vec{\psi}\|}.
\end{equation} 

The motion of the interface under the flow field $\vec{u}$ is modeled by a standard advection equation
\begin{equation}
	\label{eq:level-set-advection}
	\frac{\partial \phi}{\partial t}+\vec{u}\cdot\gphi=0.
\end{equation} 
The evolution of the gradient field is obtained by taking the gradient of the level set advection equation,
\begin{equation}
	\frac{\partial\vec{\psi}}{\partial t}+\vec{u}\cdot\nabla\vec{\psi}+\nabla\vec{u}\cdot\vec{\psi}=0.
\end{equation}
%Even if $\phi$ is signed distance function at the beginning, it normally loses this property over time due to the deformations induced by the velocity field.
%Therefore a reinitialization process is typically required after each time step.

Using the level set function one is able to define the viscosity
 at any location in the computational domain, $\vec{x}$,  with a single relation,
\begin{equation}
	\mu(\vec{x})=\mu^{-}+(\mu^{+}-\mu^{-})H(\phi(\vec{x})),
	\label{eq:ViscosityEq}
\end{equation}
where $H$ is the Heaviside function. Different forms of the Heaviside approximations have been proposed \cite{Peskin2002,Smereka2006,Engquist2005}. 
Here an accurate method 
based on an integral calculation \cite{Towers2008} is used. The Heaviside is defined as
\begin{equation}
         H(\phi(\vec{x})) = \dfrac{\nabla\Im(\phi(\vec{x}))\cdot\nabla\phi(\vec{x})}{\Arrowvert(\nabla\phi(\vec{x}))\Arrowvert^2}
\end{equation}
where
  \begin{align*}
    \Im(z) = \int_0^z \bar{H}(\zeta)d\zeta
   \qquad\text{and}\qquad
 \bar{H}(\zeta) &=
  \begin{cases}
   0        & \text{if } \zeta < 0 \\
   1        & \text{if } \zeta > 0 
  \end{cases}
  \end{align*}
Similarly the Dirac delta function is expressed as 
 \begin{equation}
   \delta(\phi(\vec{x})) = \dfrac{\nabla H(\phi(\vec{x}))\cdot\nabla\phi(\vec{x})}{\Arrowvert(\nabla\phi(\vec{x}))\Arrowvert^2}
   \label{eq:dirac}
\end{equation}
These definitions will be used to localize the contributions of interface forces in Navier-Stokes equations.

%%%%%%%%%%%%%%%%%%%%%%%%%%%%%%%%%%%%%%%%%%%%%%%%%%%%%%%%%%%%%%%%%%%%%%%%%%%%%%%%%%%%%%%%%%%%%%%%%%%%%%%%%%%%%%%%%%%%%%%%%%%
\subsection{Continuum Surface Force Model}
\label{sec:2.4}

The definitions of the Dirac and Heaviside functions in Sec. \ref{sec:2.3} allow for the use of a single equation to describe
the dynamics of the fluid over the entire domain \cite{Chang1996}. This is accomplished by 
writing the singular contributions of the bending, tension, and electric field forces as localized body force terms, 
similar to what has been done for two dimensional vesicles \cite{salac2011,salac2012}. This results in the following
single-fluid formulation,
\begin{equation}
 \begin{split}
			\hspace{-10mm} \rho\frac{D \vec{u}}{Dt}=&-\nabla p + \nabla\cdot\left(\mu\left(\nabla\vec{u}+\nabla^{T}\vec{u}\right)\right)  \\
				 & + \delta\left(\phi\right)\Arrowvert\gphi\Arrowvert\left(\nabla_s\gamma-\gamma H\gphi\right) \\
				 & + \kappa_{c}\delta(\phi)\left(\frac{H^3}{2} -2KH+\nabla^2_s H \right)\gphi\\
				  & + \delta(\phi)\Arrowvert\gphi\Arrowvert\left[\epsilon\left(\vec{E}\vec{E}-\frac{1}{2}\left(\vec{E}\cdot\vec{E}\right)\vec{I}\right) \right]\cdot\vec{n},
  \end{split}
  \label{eq:CSF_model}
\end{equation}
with
\begin{align}
		\nabla\cdot\vec{u}&=0 \hspace{3mm} \text{in} \hspace{3mm}  \Omega, \label{eq:incompressibility}\\
		\nabla_s\cdot\vec{u}&=0 \hspace{3mm} \text{on} \hspace{3mm}  \Gamma. \label{eq:inextensibility}
\end{align}
Note that in this formulation all surface quantities, such as tension or the jump in the Maxwell stress tensor, are calculated on the interface
and extended such that they are constant in the direction normal to the interface.

%%%%%%%%%%%%%%%%%%%%%%%%%%%%%%%%%%%%%%%%%%%%%%%%%%%%%%%%%%%%%%%%%%%%%%%%%%%%%%%%%%%%%%%%%%%%%%%%%%%%%%%%%%%%%%%%%%%%%%%%%%%
\subsection{Dimensionless Parameters}
\label{sec:2.5}
When an electric field is applied to the system charges from the bulk fluids migrate towards the interface.
The time scale associated with this migration process is given by the bulk charge relaxation time \cite{Saville1997,Melcher1969},
\begin{equation}
    t_c^{\pm}= \frac{\epsilon^{\pm}}{s^{\pm}}.
    \label{eq:bulk_charge_relaxation_time}
\end{equation} 

For an ion-impermeable membrane the characteristic time scale associated with the charging process is given by \cite{tsong1991,kinosita1988}
\begin{equation}
    t_m=a C_m \left(\frac{1}{s^-} + \frac{1}{2s^+}\right),
    \label{eq:relaxation_time}
\end{equation}
where $a$ is the characteristic length scale.
The applied electric stresses act to deform the vesicle membrane on the electrohydrodynamic time scale given by
\begin{equation}
	 t_{ehd}=\frac{\mu^+(1+\eta)}{\epsilon^+ E_0^2},
	\label{eq:EHD_TimeScale}
\end{equation}
where the applied electric field has a strength of $E_0$ and $\eta=\mu^-/\mu^+$ is the viscosity ratio.
The bending forces act to restore the membrane to an equilibrium configuration on the bending time scale,
\begin{equation}
	 t_{\kappa}=\frac{\mu^{+}a^3(1+\eta)}{\kappa_c}.
	\label{eq:Bending_TimeScale}
\end{equation}
Finally, the response time of a vesicle to any externally applied shear flow is simply the inverse of the shear rate, $\dot{\gamma}_0$,
\begin{equation}
	 t_{\dot{\gamma}}=\frac{1}{\dot{\gamma}_0}.
	\label{eq:shear_TimeScale}
\end{equation}

It is beneficial to estimate the order of magnitude for the times
at which different physical processes occur. Typical values of the physical properties 
have been reported as $a\approx 20\;\mu$m, $\kappa_c \approx 10^{-19}$ J, $s^+ \approx 10^{-3}$ S/m, $s^-=s^+/10$,
$\epsilon^+ \approx 10^{-9}$ F/m, $C_m \approx 10^{-2}$ F/m$^{2}$,
$\rho \approx 10^3$ kg/m$^3$, $\dot{\gamma}_0 \approx 1$ $s^{-1}$ and $\mu^-=\mu^+ \approx 10^{-3}$ Pa s \cite{Riske2006,salipante2014,needham1989electro,sadik2011vesicle}.
Using these values the bulk charge relaxation time is $t_c \approx 10^{-6}$ s, the bending time scale is $t_{\kappa_c} \approx 160$ s, 
the membrane charging time scale is $t_{m} \approx 2\times 10^{-3}$ s,
the time scale for shear is $t_{\dot{\gamma}} = 1$ s and for
an electric field of $E_0=10^{5}$ V/m the electrohydrodynamics time scale will be $t_{ehd} \approx 10^{-4}$ s.

It is worth noting that the bulk charge accumulation on the interface happens in a much faster time than any other events. 
Therefore, it can be concluded
that the electric field adjusts to a new configuration of the vesicle and fluid almost instantaneously
and the quasi-static assumption for the electric field in Eq (\ref{eq:Elec_field}) is valid. It is also important to note 
that with this parameter set the electrohydrodynamics time scale, $t_{ehd}$, is faster than the membrane charging time scale, $t_m$.
If this is not the case then the vesicle membrane will not be able to respond quickly enough to the applied forces, and only small
deformations will be observed \cite{salipante2014,schwalbe2011}.

%Different ratios of the physical time scales lead to the definitions for some key dimensionless parameters.
%Diving the bending time scale to the shear rate time scale gives a Capillary-like number, $Ca$, defined as \cite{vlahovska2009elec,schwalbe2011}
%\begin{equation}
	 %Ca = \frac{t_{\dot{\gamma}}}{t_{\kappa_c}}  =\frac{\mu^{+} \dot{\sigma_0} a^3}{\kappa_c}.
	%\label{Cadef}
%\end{equation}
%The ratio of the shear rate time scale to electrohydrodynamics time scale results in another important parameter, the Mason number $Mn$, given as \cite{schwalbe2011}
%\begin{equation}
	 %Mn = \frac{t_{\dot{\gamma}}}{t_{ehd}}  = \frac{\epsilon E_0^2}{\mu^+ \dot{\sigma_0}}.
	%\label{Mndef}
%\end{equation}

\subsection{Nondimensional Model}
\label{sec:nondim-model}
Given a characteristic length $a$ and time $t_0$ the characteristic velocity is given by $u_0=a/t_0$.
Material quantities such as viscosity and permittivity are normalized 
by their counterparts in the (outer) bulk fluid (\textit{i.e.} $\hat{\mu}=\mu/\mu^+$ and $\hat{\epsilon}=\epsilon/\epsilon^+$).
%The Reynolds number is given as .
The dimensionless fluid equations are written as
\begin{equation}
	\begin{split}
			\frac{D \vec{\hat{u}}}{D\hat{t}}=&-\hat{\nabla} \hat{p} + \frac{1}{Re}\hat{\nabla}\cdot(\hat{\mu}(\hat{\nabla}\vec{\hat{u}}+\hat{\nabla}^{T}\vec{\hat{u}}))  \\
				 & + \delta(\phi)\Arrowvert\hat{\nabla} \phi\Arrowvert\left(\hat{\nabla}_s\hat{\gamma}-\hat{\gamma} \hat{H}\hat{\nabla} \phi\right) \\
				 & + \frac{1}{Ca\;Re}\delta(\phi)\left(\frac{\hat{H}^3}{2} -2\hat{K}\hat{H}+\hat{\nabla}^2_s \hat{H} \right)\hat{\nabla} \phi\\
				  & + \frac{Mn}{Re}\delta(\phi)\Arrowvert\hat{\nabla} \phi\Arrowvert\left[\hat{\epsilon}\left(\vec{\hat{E}}\vec{\hat{E}}-\frac{1}{2}\left(\vec{\hat{E}}\cdot\vec{\hat{E}}\right)\vec{I}\right) \right]\cdot\vec{n}
  \end{split}
  \label{normalizedNS}
\end{equation}
where dimensionless quantities are denoted by a hat. 
The velocity at the boundary of the domain is given as $u_{\infty}=\chi \hat{y}$ where $\chi=\dot{\gamma}t_0$ is the
normalized applied shear rate. The uniform DC electric field at the boundary is imposed as
$\Phi_{\infty}= \hat{E} \hat{y}$ where $\hat{E}$ is the normalized strength of the applied electric field.
The dimensionless parameters are defined as follows. The Reynolds number is taken to be $Re=\rho u_0 a/\mu^+$, 
while the strength of the bending is given by a capillary-like parameter, $Ca=t_\kappa/t_0$ and the strength of the electric field
is given by the Mason number, $Mn=t_0/t_{ehd}$.

The time-evolution equation of the transmembrane potential is nondimensionalized in a similar manner:
\begin{equation}
	\hat{C}_m\frac{d\hat{V}_m}{d\hat{t}}+\hat{G}_m \hat{V}_{m}=\lambda\vec{\hat{n}}\cdot\vec{\hat{E}^{-}}=\vec{\hat{n}}\cdot\vec{\hat{E}^{+}},
	\label{eq:Vm_nondimension} 
\end{equation}
where the dimensionless membrane capacitance is given as $\hat{C}_m=\left(C_m a\right)/\left(t_0 s^+\right)$, 
the dimensionless membrane conductivity is $\hat{G}_m=\left(G_m a\right)/s^+$ and 
the conductivity ratio is expressed as  $\lambda=s^-/s^+$. 

%~ Using typical experimental parameters gives $\hat{C}_m\approx 0.1$ and $\hat{G}_m=0$, as
%~ membrane conductance is usually ignored.

For simulations in the absence of shear flow the most appropriate time scale is the membrane-charging time, $t_m$ \cite{salipante2014}. Hence for all the electrohydrodynamic
computations with no imposed shear flow ($\chi=0$) in Sec. \ref{sec:EHD}  the simulation time scale is set to $t_0=t_m$ and the dimensionless parameters $Re$, $Ca$ and $Mn$ are 
calculated accordingly.
 On the other hand, using the membrane charging time scale for the hydrodynamic simulations ($Mn=0$) will lead to abnormally huge number of iterations. 
 Therefore for the hydrodynamic computations in Sec. \ref{sec:HD} the simulation time scale is set to
 the time scale associated with shear flow ($t_0=t_{\dot{\gamma}}$). This time scale has been also used for the investigation of the vesicle dynamics under 
  combined shear flow and weak electric field in Sec. \ref{sec:Shear_Efield}. This seems to be a rational choice as the time scale of the applied weak electric field is 
  in the same range as the one for the shear flow. 
 
  Using the typical experimental values of the physical parameters 
discussed earlier in Sec. \ref{sec:2.5} the non-dimensional numbers for simulations in the absence of shear
are $t_0=t_m=2.1\times10^{-3}$ s, $\hat{C}_m=0.095$, $\hat{G}_m=0$, $Re=0.19$, $Ca=3.8\times10^{4}$, $Mn=18$, $E_0=1$,  and $\chi=0$.
Clearly, at this time scale the Reynolds number is not small enough to justify the 
Stokes approximation while the bending contribution is almost negligible. Despite the small contribution from the bending 
in this situation it is kept in the formulation as it may become important during other time-scales, such as when shear is applied.

For simplicity the hat notation in the equations henceforth dropped. 
All the above mentioned parameters along with the viscosity ratio, $\eta$, and reduced volume, $v$, will be used
to investigate the dynamics of a three-dimensional vesicle in different flow conditions and strength of electric field.

%%%%%%%%%%%%%%%%%%%%%%%%%%%%%%%%%%%%%%%%%%%%%%%%%%%%%%%%%%%%%%%%%%%%%%%%%%%%%%%%%%%%%%%%%%%%%%%%%%%%%%%%%%%%%%%%%%%%%%%%%%%
\section{Numerical Methodology}
\label{sec:3.0}

In this section the specific numerical implementation of the non-dimensional evolution equations described in Sec. \ref{sec:nondim-model} will be given.

%%%%%%%%%%%%%%%%%%%%%%%%%%%%%%%%%%%%%%%%%%%%%%%%%%%%%%%%%%%%%%%%%%%%%%%%%%%%%%%%%%%%%%%%%%%%%%%%%%%%%%%%%%%%%%%%%%%%%%%%%%%
 \subsection{Electric field solution: Immersed Interface Method}
\label{sec:3.1}

Using a second order time discretization scheme for the time-evolution equation of the transmembrane potential in (\ref{eq:Vm_nondimension})
and treating the term $V_{m}$ implicitly one obtains
\begin{equation}
	\hat{C}_m\frac{3{V}^{n+1}_m- 4{V}^{n}_m+ {V}^{n-1}_m}{2\Delta t}+\hat{G}_m {V}^{n+1}_{m}=\lambda\vec{n}\cdot\vec{E^{-}}=\vec{n}\cdot\vec{E^{+}}.
	\label{eq:Vm_discretization} 
\end{equation}
The right hand side of Eq. (\ref{eq:Vm_discretization}) shows that the solution of the electric field at the current time step, $t^{n+1}$, is needed in order to compute
 $V^{n+1}_m$. To solve the electric field equation with mismatched conductivity, a three-dimensional Immersed Interface Method (IIM)
with time-varying solution jumps is employed. 
To achieve second-order accuracy in space, the jump conditions for the electric potential are derived up to 
the second normal derivative. First define $r=-\vec{n}\cdot\vec{E}^{\alpha}$ with $\alpha$ representing the domain with lower 
conductivity (\textit{i.e.} if $s^-<s^+$ then $\alpha=-$). Using this definition and employing the relations given in Eqs. (\ref{eq:Phi_Difference}), (\ref{eq:current_denisity_continuity}) 
and (\ref{eq:Vm_discretization}) the complete set of jump conditions can be derived as (for details of the derivation see \cite{Kolahdouz2013})
\begin{align}
    \left[\Phi\right]&=\frac{0.5V^{n-1}_m-2V^{n}_m + \Delta t s^{\alpha}r}{1.5\hat{C}_m +\Delta t \hat{G}_m} \label{Phi_Jump_r}\\
     \left[\frac{\partial\Phi}{\partial n}\right]&=-\frac{[s]}{s^{-\alpha}}r, \label{PhiPrime_Jump_r} \\
    \left[\frac{\partial^2\Phi}{\partial n^2}\right]&=\nabla^2_s\biggl(\frac{2V^{n}_m-0.5V^{n-1}_m - \Delta t s^{\alpha}r}{1.5\hat{C}_m +\Delta t \hat{G}_m}\biggl)+H\frac{[s]}{s^{-\alpha}}r, \label{PhiDoublePrime_r} 
\end{align}
where $s^{-\alpha}$ represents the higher of the two fluid conductivities.

The field equation Eq. (\ref{eq:Elec_lap}), along with the above set of jump conditions are used to solve for $r$ and $\Phi$ simultaneously. The new 
trans-membrane potential is then updated using Eq. (\ref{eq:Vm_discretization}), noting that $r$ provides the solution to the electric field in one of the domains.
The technique to solve the 
resulting linear system plus further details of the implementation and sample results can be found in Ref. \cite{Kolahdouz2013}.

To calculate the electric field contribution to the Navier-Stokes equation, Eq. (\ref{normalizedNS}),
the normal component of the Maxwell stress tensor is needed. This tensor is computed at the interface for each domain,
using the corrections in jump conditions to obtain second-order accuracy. The difference between the normal component is then calculated, giving the electric field force on the interface.

%%%%%%%%%%%%%%%%%%%%%%%%%%%%%%%%%%%%%%%%%%%%%%%%%%%%%%%%%%%%%%%%%%%%%%%%%%%%%%%%%%%%%%%%%%%%%%%%%%%%%%%%%%%%%%%%%%%%%%%%%%%
 \subsection{Level Set Advection}
\label{sec:3.2}
Many of the equations given in previous sections require high order geometric quantities, such as the curvature. To ensure that surface quantities are 
smooth a semi-implicit level set advection scheme is utilized. The method used here is an improvement of the semi-implicit, gradient augmented
level set \cite{Kolahdouz20132}, and is briefly described here.

\begin{figure}[ht!]
        \begin{center}
            \subfigure[$Advection$] {
                  \includegraphics[width=0.3\textwidth]{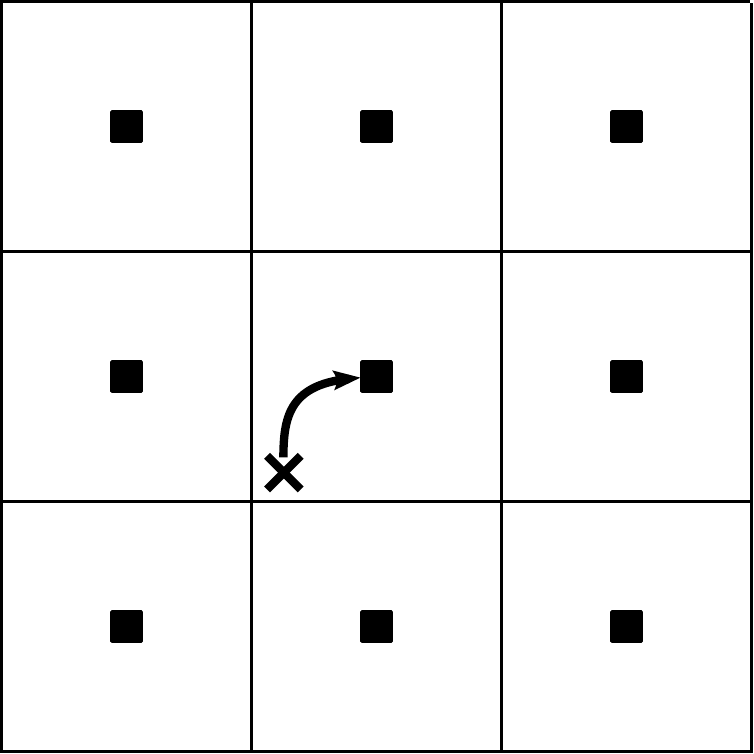}
                \label{fig:Advection}
            }             
            \subfigure[$Smoothing$]{
                  \includegraphics[width=0.3\textwidth]{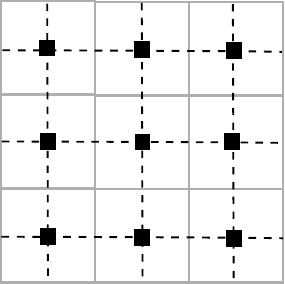}
                \label{fig:Smoothing}
            }
            \subfigure[$Interpolation$] {
                  \includegraphics[width=0.3\textwidth]{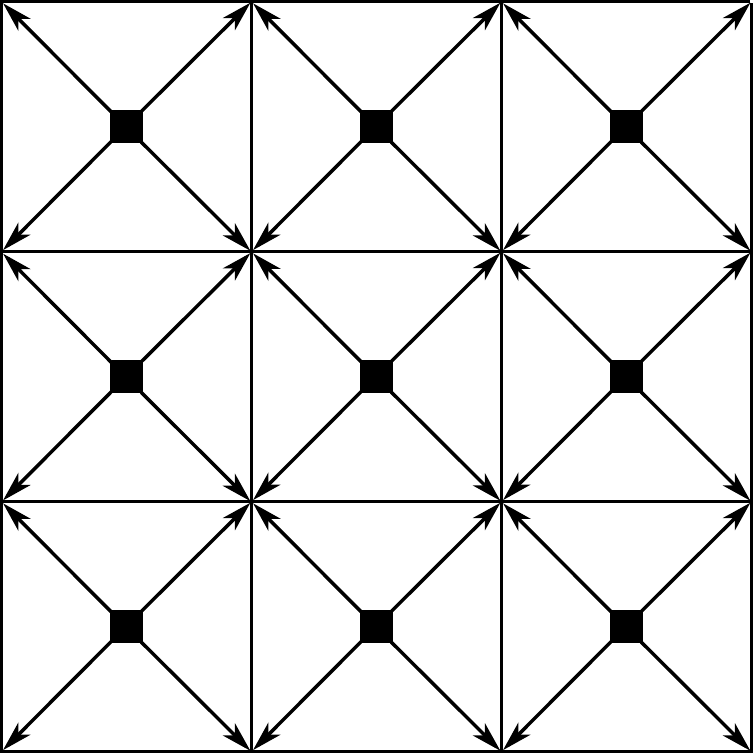}
                \label{fig:Interpolation}
			}
			\label{fig:graphicSemiJet}
            \caption{ Schematic representation of the steps in Semi-implicit Jet Schemes.
            }    
            \label{fig:volume_conserving}
        \end{center}
        
    \end{figure}

Consider the advection of the level set field, $\phi$, and the associated gradient field, $\vec{\psi}$, due to an underlying flow field, $\vec{u}$. 
Seibold \textit{et. al.} generalized this type of advection by considering a sub-grid of points centered on a grid point, which 
provides a \textit{jet} of level set information \cite{Seibold2012}. It was shown that using a sub-grid spacing of $\mathcal{O}(\delta^{1/4})$,
where $\delta$ is the floating point operation accuracy, provided the optimal accuracy for linear advection equations. As the equations which model
the electrohydrodynamics of vesicles require that a smooth level set field be maintained at all times, this high accuracy solution is relaxed.
In this work the jet of level-set information surrounding a grid point is the cell centers, Fig. \ref{fig:Advection}.

The advection of the level set jet proceeds with the following three steps: 
\begin{inparaenum}[1)]
	\item advection of cell-centers,
	\item smoothing at cell-centers, and
	\item projection onto grid points.
\end{inparaenum}
Unlike the original semi-implicit, gradient-augmented level set method \cite{Kolahdouz20132} this has the advantage of only requiring a single smoothing operation,
as described below.

The advection of cell-centers proceeds by using a Lagrangian advection scheme:
\begin{align}
	\vec{x}_d &= \vec{x}_c - \Delta t \vec{u}_c^n, \\
	\hat{\phi}^c &= I_3\left(\phi^n, \vec{x}_d\right),
\end{align}
where $\vec{x}^c$ is the cell-center location, $\hat{\phi}^c$ is the tentative level set value at the cell center, and $I_3\left(\phi^n,\vec{x}_d\right)$ is a cubic interpolant
of the previous time-steps level set values evaluated at the departure location, $\vec{x}_d$.

The values at cell-centers are then smoothed using a semi-implicit technique first introduced for standard level set advection \cite{Smereka2003}:
\begin{equation}
	\frac{\phi^c-\hat{\phi}^c}{\Delta t}=\beta\nabla^2\phi^c-\beta\nabla^2\hat{\phi}^c,
\end{equation}
where $\beta$ is a user-parameter, typically chosen to be 0.5 \cite{Kolahdouz20132, Smereka2003}.
Note that the grid in this smoothing step is the cell-center (offset) grid. The boundary condition is the continuity of derivatives using the tentative
level set values. For example, on the right edge the boundary condition is
\begin{equation}
	\left(\frac{\partial \phi^c}{\partial x}\right)_{N_x, j, k} = \frac{\hat{\phi}^c_{N_x-1/2,j,k}-\hat{\phi}^c_{N_x-3/2,j,k}}{h_x},
\end{equation}
where $N_x$ and $h_x$ are the the number of grid points and the grid-spacing in the $x$-direction, respectively.

Finally, the cell-center values are projected back onto the grid points by using second-order stencils. For example, the updated level set values at grid point $(i,j,k)$ are given by
\begin{align}
	\phi_{i,j,k}=&\frac{1}{8}\left(\phi^c_{i+1/2,j+1/2,k+1/2} + \phi^c_{i-1/2,j+1/2,k+1/2}+\phi^c_{i+1/2,j-1/2,k+1/2}\right.\\ \nonumber
			&+\phi^c_{i-1/2,j-1/2,k+1/2}+\phi^c_{i+1/2,j+1/2,k-1/2} + \phi^c_{i-1/2,j+1/2,k-1/2}\\ \nonumber
			&\left.+\phi^c_{i+1/2,j-1/2,k-1/2}+\phi^c_{i-1/2,j-1/2,k-1/2}\right),
\end{align}
while the gradient in the $x$-direction is given by
\begin{align}
	\psi^x_{i,j,k}=&\frac{1}{8h_x}\left(\phi^c_{i+1/2,j+1/2,k+1/2} - \phi^c_{i-1/2,j+1/2,k+1/2}+\phi^c_{i+1/2,j-1/2,k+1/2}\right.\\ \nonumber
			&-\phi^c_{i-1/2,j-1/2,k+1/2}+\phi^c_{i+1/2,j+1/2,k-1/2} - \phi^c_{i-1/2,j+1/2,k-1/2}\\ \nonumber
			&\left.+\phi^c_{i+1/2,j-1/2,k-1/2}-\phi^c_{i-1/2,j-1/2,k-1/2}\right).
\end{align}
The other gradient values can be similarly evaluated.
The advantage of this method over that of Ref. \cite{Kolahdouz20132} is that only a single smoothing operation is required, versus one for the level set and 
three for the gradient fields. Work is currently underway investigating the full properties of this semi-implicit jet scheme.

\subsection{Level Set Reinitialization and Closest Point Calculation}
\label{sec:lsReinit}
After the advection of the level set a reinitialization procedure is used. This is done for two reasons. First, it has been 
shown that reinitialization aids in the conservation of mass in level set-based simulations \cite{Sussman1999}. Second, the 
point on the interface with the shortest distance to a given grid point, called the closest-point, is required for the evaluation of the surface
equations given below. The method employed here uses
a variant of Newton's method and a tricubic interpolation to find the closest point on the interface from an arbitrary grid location in the neighborhood, 
see Ref. \cite{chopp2001} for more details. 
Once the closest points are determined the level set function is replaced by the corresponding signed distance value while the gradient field is set to the unit outward normal vector, which is
the normalized vector pointing from the closest point to the grid point, corrected so that it points outward.
This procedure does not need to be done everywhere in the domain, but only near the interface. In the results below all nodes within five grid points of the interface have their closest point 
calculated and explicitly reinitialized. The remaining nodes are reinitialized using a first-order PDE based reinitialization scheme \cite{Peng1999410,Russo200051}.

%%%%%%%%%%%%%%%%%%%%%%%%%%%%%%%%%%%%%%%%%%%%%%%%%%%%%%%%%%%%%%%%%%%%%%%%%%%%%%%%%%%%%%%%%%%%%%%%%%%%%%%%%%%%%%%%%%%%%%%%%%%
 \subsection{Hydrodynamics Solution: Projection Method}
\label{sec:3.3}

There are four total conservation conditions that must be satisfied during the course of the simulation:
\begin{inparaenum}[1)]
	\item local surface area,
	\item total surface area,
	\item local fluid volume,
	\item total fluid volume.
\end{inparaenum}
Vesicle dynamics are extremely sensitive to any changes in these quantities \cite{salac2011,salac2012,biben2005,Biben2011} and thus high accuracy is required. Unfortunately, 
error in the solution of the fluid equations, in addition to the non-conservation properties of the level set method, can induce 
large errors over the course of the simulation. To account for this, the fluid equations will be modified to
explicitly correct any error accumulation. The idea presented here is based on the work of Laadhari \textit{et. al.} \cite{laadhari2012}.
The difference is that in the previous work the corrections were not reflected in the fluid field, but instead modified the velocity used to advect the level set.
Here, the corrections are included in the fluid field calculation and therefore a modified advection field is not required.
Preliminary results for standard multiphase flow problems were presented in Ref. \cite{salac2015} and are extended here for vesicles.

A projection method is implemented to solve for the velocity, 
pressure and tension. 
First, a semi-implicit update is performed to obtain a tentative velocity field,
\begin{equation}
	\dfrac{\vec{u}^{\ast}-\vec{u}^n}{\Delta t}+\vec{u}^n\cdot\nabla\vec{u}^{\ast}=-\nabla p^n +\dfrac{1}{Re}\nabla\cdot\left(\nabla \vec{u}^{\ast}+\left(\nabla\vec{u}^n\right)^T\right)+\vec{f}_H^n+\vec{f}_\gamma^n+\vec{f}_{el}^n,
\end{equation}
where $\vec{f}_H$, $\vec{f}_\gamma$ and $\vec{f}_{el}$ are the bending, tension, and electric field forces localized around the interface, Eq. (\ref{eq:CSF_model}).    
The superscript $n$ refers to the solution at the previous time step.

Next, the tentative velocity field is projected onto the divergence and surface-divergence free velocity space,
\begin{equation}
	\dfrac{\vec{u}^{n+1}-\vec{u}^{\ast}}{\Delta t}=-\nabla q +\delta(\phi)\|\nabla\phi\|\left(\nabla_s \xi-\xi H\nabla\phi \right),
	\label{eq:projection}
\end{equation}
where $q$ and $\xi$ are the corrections needed for the pressure and tension, respectively.
Finally, the pressure and tension are updated by including the corrections,
\begin{align}
	p^{n+1}&=p^n+q,\\
	\gamma^{n+1}&=\gamma^n+\xi.
\end{align}

The four conservation conditions can be written as \cite{laadhari2012}
\begin{align}
	\nabla\cdot\vec{u}^{n+1}&=0 \qquad &\textnormal{(local volume conservation),} \\
	\int_\Gamma\vec{n}\cdot\vec{u}^{n+1}\;dA&=\frac{dV}{dt} \qquad &\textnormal{(global volume conservation),} \\
	\nabla_s\cdot\vec{u}^{n+1}&=0 \qquad &\textnormal{(local area conservation),} \\
	\int_\Gamma H \vec{n}\cdot\vec{u}^{n+1}\;dA&=\frac{dA}{dt} \qquad &\textnormal{(global area conservation).} 
\end{align}
The use of only the pressure and tension is not sufficient to satisfy all four conservation conditions.
There, the pressure and tension fields are split into a constant and spatially-varying component:
\begin{align}
	p &= \tilde{p} + (1-H(\phi))p_0, \\
	\gamma &= \tilde{\gamma} + \gamma_0,
\end{align}
where $\tilde{p}$ and $\tilde{\gamma}$ are spatially varying while $p_0$ and $\gamma_0$ are constant. Note that $\tilde{p}$, $\tilde{\gamma}$,
$p_0$, and $\gamma_0$ all vary in time. Conceptually, this splitting allows for the enforcement of local conservation
through $\tilde{p}$ and $\tilde{\gamma}$ while global conservation is enforced through $p_0$ and $\gamma_0$.

The corresponding corrections are now $q=\tilde{q}+(1-H(\phi))q_0$, and $\xi=\tilde{\xi}+\xi_0$, while the projection 
step, Eq. (\ref{eq:projection}), is now written as
\begin{equation}
	\vec{u}^{n+1}=\vec{u}^{\ast} + \Delta t\left(-\nabla \tilde{q} + \delta(\phi)q_0 \nabla \phi + \delta(\phi)\|\nabla\phi\|\left(\nabla_s\tilde{\xi}-\tilde{\xi} H \nabla \phi-\xi_0 H \nabla\phi\right)\right).
	\label{eq:expandedProjection}
\end{equation}

Noting that the time derivatives of the volume and area are to correct any accumulated errors in the solution,
and using Eq. (\ref{eq:expandedProjection}), the four conservation equations can be written in terms of the 
four unknowns ($\tilde{q}$, $\tilde{\xi}$, $q_0$, and $\xi_0$), the current area and volume, and the initial area and volume. Specifically, applying the local area conservation
equation requires that
\begin{equation}
	-\nabla\cdot\vec{u}^\ast = \Delta t\nabla\cdot\left(-\nabla \tilde{q} + \delta(\phi)q_0 \nabla \phi + \delta(\phi)\|\nabla\phi\|\left(\nabla_s\tilde{\xi}-\tilde{\xi} H \nabla \phi+\xi_0 H \nabla\phi\right)\right),
	\label{eq:localVolume}
\end{equation}
while the total volume conservation requires that
\begin{align}
	\frac{V^0-V^n}{\Delta t}-&\int_\Gamma\vec{n}\cdot\vec{u}^{\ast}\;dA = \nonumber\\ 
		&\Delta t\int_\Gamma\left(-\vec{n}\cdot\nabla \tilde{q} + \delta(\phi)q_0 \|\nabla \phi\| - \delta(\phi)\|\nabla\phi\|^2\left(\tilde{\xi} H +\xi_0 H \right)\right)\;dA,
	\label{eq:globalVolume}
\end{align}
where $V^n$ is the current volume and the time-derivative of the volume is chosen so that at the end of the time-step the volume equals the initial volume, $V^0$.

Conservation of local and global area results in the following two equations:
\begin{equation}
	-\nabla_s\cdot\vec{u}^{\ast} = \Delta t\nabla_s\cdot\left(-\nabla \tilde{q} + \delta(\phi)q_0 \nabla \phi + \delta(\phi)\|\nabla\phi\|\left(\nabla_s\tilde{\xi}-\tilde{\xi} H \nabla \phi-\xi_0 H \nabla\phi\right)\right),
	\label{eq:localArea}
\end{equation}
and 
\begin{align}
	\frac{A^0-A^n}{\Delta t}-&\int_\Gamma H\vec{n}\cdot\vec{u}^{\ast}\;dA = \nonumber\\ 
		&\Delta t\int_\Gamma H\left(-\vec{n}\cdot\nabla \tilde{q} + \delta(\phi)q_0 \|\nabla \phi\| - \delta(\phi)\|\nabla\phi\|^2\left(\tilde{\xi} H +\xi_0 H \right)\right)\;dA,
	\label{eq:globalArea}
\end{align}
where $A^0$ is the initial surface area and $A^n$ is the current surface area.
	
Let $\vec{\tilde{q}}$ represent the vector holding the values of $\tilde{q}$ in the entire, discretized domain and $\vec{\tilde{\xi}}$ is the vector holding
the values of $\tilde{\xi}$ in the entire, discretized domain. It is now possible to represent the conservation relationships,
Eqs. (\ref{eq:localVolume})-(\ref{eq:globalArea}), as linear operators acting on $\vec{\tilde{q}}$, $q_0$, $\vec{\tilde{\xi}}$, and $\xi_0$. 
Local volume conservation, Eq. (\ref{eq:localVolume}),
can be represented as
\begin{equation}
	\vec{L}\vec{\tilde{q}}+\vec{l} q_0+\vec{L}_{\xi}\vec{\tilde{\xi}}+\vec{l}_{\xi}\xi_0=-\vec{D}\vec{u}^\ast,
\end{equation}
where $\vec{D}$ is the discrete divergence operator and
\begin{align}
	\vec{L}\vec{\tilde{q}}\approx&-\Delta t\nabla\cdot\nabla\tilde{q},  &\vec{l} q_0\approx q_0 \Delta t\nabla\cdot\left(\delta(\phi)\nabla\phi\right),\\
	\vec{L}_{\xi}\vec{\tilde{\xi}}\approx&\Delta t\nabla\cdot\left(\delta(\phi)\|\nabla\phi\|\left(\nabla_s\tilde{\xi}-\tilde{\xi}H\nabla\phi\right)\right),  &\vec{l}_{\xi}\xi_0\approx \xi_0\Delta t\nabla\cdot\left(\delta(\phi)\|\nabla \phi\| H\nabla\phi\right),
\end{align}
are the discretizations of the continuous operations. Note that $\vec{L}$ and $\vec{L}^{\xi}$ are a linear operator matrices while $\vec{l}$ and $\vec{l}_{\xi}$ are 
linear operator vectors.

The integrals over the vesicle interface can be approximated as summations over the discretized domain: $\int_\Gamma f\;dA\approx \sum\delta_{i,j,k}f_{i,j,k}\Delta V$,
where $\delta_{i,j,k}$ is the Dirac delta function defined in Eq. (\ref{eq:dirac}) at a grid point, $f_{i,j,k}$ is the function value at that grid point, and 
$\Delta V=h_x h_y h_z$ is the volume of a cell. In linear operator form this calculated by taking the dot product between the integration vector and the vector containing
the function values. Define the following linear operations:
\begin{align}
	\vec{s}^T\vec{\tilde{q}}\approx& -\Delta t\int_\Gamma\left(\vec{n}\cdot\nabla\tilde{q}\right)dA, &a q_0\approx&q_0\Delta t\int_\Gamma\delta(\phi)\|\nabla\phi\|dA, \\
	\vec{s}_\xi^T\vec{\tilde{\xi}}\approx&-\Delta t\int_\Gamma\left(\delta(\phi)\|\nabla\phi\|^2\tilde{\xi}H\right)dA, 	&a_\xi \xi_0\approx&\xi_0 \Delta t\int_\Gamma\left(\delta(\phi)\|\nabla\phi\|^2H\right)dA.
\end{align}
In this case $\vec{s}$ and $\vec{s}_\xi$ are linear operator vectors while $a$ and $a_\xi$ are scalar values. This results in the following linear equation:
\begin{equation}
	\vec{s}^T\vec{\tilde{q}}+a q_0+\vec{s}_{\xi}^T\vec{\tilde{\xi}}+a_{\xi}\xi_0=E_V,
\end{equation}
where $E_V\approx(V^0-V^n)/\Delta t-\int_\Gamma\left(\vec{n}\cdot\vec{u}^\ast\right)dA$ is the discrete form of the global volume correction needed.

The surface-conservation equations are written in a similar manner:
\begin{equation}
	\vec{L}^s\vec{\tilde{q}}+\vec{l}^s q_0+\vec{L}^s_{\xi}\vec{\tilde{\xi}}+\vec{l}^s_{\xi}\xi_0=-\vec{D}_s\vec{u}^\ast,
\end{equation}
where $\vec{D}_s$ is the discrete surface-divergence operator and
\begin{align}
	\vec{L}^s\vec{\tilde{q}}\approx&-\Delta t\nabla_s\cdot\nabla\tilde{q},  &\vec{l}^s q_0\approx q_0 \Delta t\nabla_s\cdot\left(\delta(\phi)\nabla\phi\right),\\
	\vec{L}^s_{\xi}\vec{\tilde{\xi}}\approx&\Delta t\nabla_s\cdot\left(\delta(\phi)\|\nabla\phi\|\left(\nabla_s\tilde{\xi}-\tilde{\xi}H\nabla\phi\right)\right),  &\vec{l}^s_{\xi}\xi_0\approx \xi_0\Delta t\nabla_s\cdot\left(\delta(\phi)\|\nabla \phi\| H\nabla\phi\right)
\end{align}
for the local surface-area conservation, Eq. (\ref{eq:localArea}) and
\begin{equation}
	\vec{s}_H^T\vec{\tilde{q}}+b q_0+\vec{s}_{H\xi}^T\vec{\tilde{\xi}}+b_{\xi}\xi_0=E_A,
\end{equation}
where $E_A\approx(A_0-A^n)/\Delta t-\int_\Gamma\left(H\vec{n}\cdot\vec{u}^\ast\right)dA$ is the discrete form of the global area correction needed and
\begin{align}
	\vec{s}_H^T\vec{\tilde{q}}\approx& -\Delta t\int_\Gamma\left(H\vec{n}\cdot\nabla\tilde{q}\right)dA, &b q_0\approx&q_0\Delta t\int_\Gamma H\delta(\phi)\|\nabla\phi\|dA, \\
	\vec{s}_{H\xi}^T\vec{\tilde{\xi}}\approx&-\Delta t\int_\Gamma\left(\delta(\phi)\|\nabla\phi\|^2\tilde{\xi}H^2\right)dA, 	&b_\xi \xi_0\approx&\xi_0 \Delta t\int_\Gamma\left(\delta(\phi)\|\nabla\phi\|^2H^2\right)dA.
\end{align}

Using the notation above the set of four linear equations can be written in matrix-vector form:
\begin{equation}
		\left[\begin{array}{cccc}
			\vec{L} 	& \vec{l} 	& \vec{L}_\xi		& \vec{l}_\xi \\
			\vec{s}^T & a 			& \vec{s}_\xi^T 	& a_\xi \\
			\vec{L}^s 	& \vec{l}^s 	& \vec{L}^s_\xi		& \vec{l}^s_\xi \\
			\vec{s}_H^T & b 			& \vec{s}_{H\xi}^T 	& b_{\xi}
		\end{array} \right]	
		\left[\begin{array}{c}
			\vec{\tilde{q}} \\
			q_0 \\
			\vec{\tilde{\xi}}\\
			\xi_0
		\end{array} \right] =
		\left[\begin{array}{c}
			-\vec{D}\vec{u}^{\ast} \\
			E_V \\
			-\vec{D}_s\vec{u}^{\ast} \\
			E_A
		\end{array} \right],
		\label{eq:blockSystem}
\end{equation}
Note that the specific spatial discretization have not yet been specified, as the general concepts are discretization-independent.

While it is possible to form a globally-assembled matrix for Eq. (\ref{eq:blockSystem}), it would be computationally expensive.
Not only would the size of the system be large, but it would also have to be re-formed every time step as many of the
components, such as $\vec{L}_{\xi}^s$, $\vec{l}_\xi^s$ and $\vec{s}_H^T$ depend on the location of the interface, which changes over time.
Instead, write the complete system in the following simplified notation,
\begin{equation}
		\left[\begin{array}{cc}
			\vec{\mathbb{A}} & \vec{\mathbb{B}} \\
			\vec{\mathbb{C}} & \vec{\mathbb{D}} 
		\end{array} \right]	
		\left[\begin{array}{c}
			\vec{q} \\
			\vec{\xi}
		\end{array} \right] =
		\left[\begin{array}{c}
			\vec{e}_V \\
			\vec{e}_A
		\end{array} \right],
		\label{eq:partMat}
\end{equation}
where 
\begin{align}
	\vec{\mathbb{A}}&=
	\left[\begin{array}{cc}
			\vec{L} 	& \vec{l} 	\\
			\vec{s}^T & a 						
		\end{array} \right], \textrm{    }\qquad
	\vec{\mathbb{B}}=
	\left[\begin{array}{cc}
			\vec{L}_\xi		& \vec{l}_\xi \\
			\vec{s}_\xi^T 	& a_\xi
		\end{array} \right], \nonumber \\ 
	\vec{\mathbb{C}}&=
	\left[\begin{array}{cc}			
			\vec{L}^s 	& \vec{l}^s\\
			\vec{s}_H^T & b 		
		\end{array} \right], \textrm{ and}\quad 
	\vec{\mathbb{D}}=
	\left[\begin{array}{cc}			
			\vec{L}^s_\xi		& \vec{l}^s_\xi \\
			\vec{s}_{H\xi}^T 	& b_{\xi}
		\end{array} \right],
\end{align}
with the combined vectors $\vec{q}=[\tilde{\vec{q}}, q_0]^T$, $\vec{\xi}=[\tilde{\vec{\xi}}, \xi_0]^T$, 
		$\vec{e}_V=[-\vec{D}\vec{u}^\ast, E_V]^T$ and $\vec{e}_A=[-\vec{D}_s\vec{u}^\ast, E_A]^T$.
Using a Schur Complement approach the solution is
\begin{equation}
		\left[\begin{array}{c}
			\vec{q} \\
			\vec{\xi}
		\end{array} \right] = 				
		\left[\begin{array}{cc}
			\vec{\mathbb{I}}_q & \vec{0} \\
			-\vec{\mathbb{D}}^{-1}\vec{\mathbb{C}} & \vec{\mathbb{I}}_\xi
		\end{array} \right]
		\left[\begin{array}{cc}
			\vec{\mathbb{S}}^{-1} & \vec{0} \\
			\vec{0} & \vec{\mathbb{D}}^{-1} 
		\end{array} \right]	
		\left[\begin{array}{cc}
			\vec{\mathbb{I}}_q & -\vec{\mathbb{B}}\vec{\mathbb{D}}^{-1} \\
			\vec{0} & \vec{\mathbb{I}}_\xi
		\end{array} \right]			
		\left[\begin{array}{c}
			\vec{e}_V \\
			\vec{e}_A
		\end{array} \right],
\end{equation}
where $\vec{\mathbb{S}}$ is the Schur Complement of the original partitioned matrix and is give by
\begin{equation}
	\vec{\mathbb{S}}=\vec{\mathbb{A}} - \vec{\mathbb{B}}\vec{\mathbb{D}}^{-1}\vec{\mathbb{C}}.
\end{equation}
The application of the Schur Complement is accomplished by a matrix-free iterative solver, such as GMRES.

Let the focus now turn to the calculation of $\mathbb{D}^{-1}$. This inverse is given implicitly through the solution 
of the following, generalized linear system:
\begin{equation}
	\left[\begin{array}{cc}
		\vec{L}_{\xi}^s & \vec{l}_{\xi}^s \\
		\vec{s}_H^T & b_\xi
	\end{array}\right]
	\left[\begin{array}{c}
		\vec{x}_0\\
		x_1
	\end{array}\right]=
	\left[\begin{array}{c}
		\vec{y}_0\\
		y_1
	\end{array}\right].
\end{equation}
Turning to the Schur Decomposition the solution can be written as
\begin{equation}
	\left[\begin{array}{c}
		\vec{x}_0\\
		x_1
	\end{array}\right] = 
	\left[\begin{array}{cc}
		\vec{I} & 0 \\
		b_\xi^{-1}\vec{s}_H^T & 1
	\end{array}\right]
	\left[\begin{array}{cc}
		\vec{S}^{-1} & 0 \\
		0 & b_\xi^{-1}
	\end{array}\right]
	\left[\begin{array}{cc}
		\vec{I} & -b_\xi^{-1}\vec{l}_{\xi}^s \\
		0 & 1
	\end{array}\right]
	\left[\begin{array}{c}
		\vec{y}_0\\
		y_1
	\end{array}\right],
\end{equation}
where the Schur Complement is a rank-1 update on the matrix $\vec{L}_{\xi}^s$:
\begin{equation}
	\vec{S} = \vec{L}_{\xi}^s-\frac{1}{b_\xi}\vec{l}_{\xi}^s\vec{s}_H^T.
\end{equation}
Using the Sherman--Morrison formula the inverse of this Schur Complement
is given by
\begin{equation}
	\vec{S}^{-1}=\left(\vec{L}_{\xi}^s\right)^{-1}+\dfrac{\left(\vec{L}_{\xi}^s\right)^{-1}\vec{l}_{\xi}^s\vec{s}_H^T\left(\vec{L}_\xi^s\right)^{-1}}{b_\xi-\vec{s}_H^T\left(\vec{L}_{\xi}^s\right)^{-1}\vec{l}_{\xi}^s}.
\end{equation}
So long as $b_\xi\neq \vec{s}_H^T\left(\vec{L}_\xi^s\right)^{-1}\vec{l}_{\xi}^s$ this inverse is defined. While this inequality will not be proven here, a check during the 
course of the simulations presented below is performed every time step and this condition has never been violated.

\subsection{Evaluation and Advection of Surface Quantities}
\label{sec:surfaceEq}
   A word must be said about the solution of surface equations. In the projection method the pressure is defined everywhere in the domain and 
thus standard techniques can be used. The tension, on the other hand, is only defined on the interface and must be handled using special methods.
Unlike the immersed boundary method or other front-tracking techniques Lagrangian points are not tracked over the course of the simulation.
Instead, all surface equations and quantities, such as the tension and trans-membrane potential, are evaluated using the Closest Point Method \cite{MacDonald2010,MacDonald2008}.
The general idea is to replace surface equations by localized equations in the embedding space by extending quantities from the interface in a systematic way. 
This is done by replacing the value at a grid point in a finite difference approximation by it's interpolated value at the closest point on the interface.
Additional information can be found in References \cite{MacDonald2010,MacDonald2008}.
	All surface quantities, such as the tension and trans-membrane potential, are advected along side the level set. 
After the level set is advected surface quantities are advected by using the total-derivative form of the advection equation for a material 
quantity $f$: 
\begin{align}
	\vec{x}_d=&\vec{x}-\Delta t\vec{u}^n, \\
	\hat{f}=&I_3\left(f^n,\vec{x}_d\right),
\end{align}
where $I_3\left(f^n, \vec{x}_d\right)$ is the evaluation of the cubic interpolant using the previous time-steps value evaluated
at the departure location $\vec{x}_d$. The values are then replaced by the values at the closest point to a grid point,
\begin{equation}
	f(\vec{x})=I_3\left(\hat{f},\vec{x}_{cp}\right),
\end{equation}
where $\vec{x}_{cp}$ is the closest point on the interface to grid point $\vec{x}$.

%%%%%%%%%%%%%%%%%%%%%%%%%%%%%%%%%%%%%%%%%%%%%%%%%%%%%%%%%%%%%%%%%%%%%%%%%%%%%%%%%%%%%%%%%%%%%%%%%%%%%%%%%%%%%%%%%%%%%%%%%%%
\subsection{Overall algorithm}
	The overall algorithm to solve the electrohydrodynamics of the vesicle is performed using a time-staggered approach:
	\\
\begin{enumerate}[Step I:]
	\item{ 
		Solve for the updated electric field and trans-membrane potential using Sec. \ref{sec:3.1}.
	}
	\item{ 
		Solve the hydrodynamics equations and obtain the updated velocity field, pressure and the tension using Sec. \ref{sec:3.3}.
	}
	\item{ 
		Using the updated velocity field from Step II advance the level set (interface) using Sec. \ref{sec:3.2}.
	}
	\item{ 
		Reinitialize the level set values around the interface and recalculate the closest point data, Sec. \ref{sec:lsReinit}.
	}
	\item{
		Advect surface quantities such as tension and trans-membrane potential using Sec. \ref{sec:surfaceEq}.\
		}
	\item{ 
		One time step has been completed. Return to Step I and repeat until final time is reached.
	}
\end{enumerate}

\label{sec:3.4}
%%%%%%%%%%%%%%%%%%%%%%%%%%%%%%%%%%%%%%%%%%%%%%%%%%%%%%%%%%%%%%%%%%%%%%%%%%%%%%%%%%%%%%%%%%%%%%%%%%%%%%%%%%%%%%%%%%%%%%%%%%%
\section{Results}
\label{sec:4.0}

In this section the dynamics of the vesicle is studied in pure shear flow, in the presence of an electric field, and in combination of the two effects.
The vesicle surface area is fixed to $4\pi$ in all the situations while the enclosed volume is varied.
For all results presented here a collocated, Cartesian mesh with uniform grid spacing in each direction is used. Periodicity is assumed in the 
$x$- and $z$-directions while wall boundary conditions are given in the $y$-direction. Unless otherwise stated the domain
is a box covering the domain $[-4.5,4.5]^3$. Any externally applied shear flow is given by imposing a velocity of $\vec{u}_{bc}=(\chi y,0,0)$ on the
wall boundaries, where $\chi$ is the normalized shear rate. Electric fields are applied by fixing the electric potential in the $y$-direction at 
$E_{bc}=E_0 y$, where $E_0$ is the normalized electric field strength. It is assumed that the trans-membrane conductance is zero and thus $\hat{G}_m=0$.
Finally, compact finite difference approximations
for all spatial derivatives have been implemented (\textit{e.g.} the pressure-Poisson term is approximated by $\nabla\cdot\nabla q\approx \nabla^2 q$ 
at the discrete level). For all results surface area and volume are all conserved to within $0.01\%$ error.

\subsection{Hydrodynamics of Vesicle ($Mn=0$)}
\label{sec:HD}

%To study the effects of different parameters, we wish to fix the dimensionless parameters in the electrohydrodynamics equation and vary the normalized boundary
%values of $\dot{\gamma}$ and $E$ in  $E_{\infty}$ and $u_{\infty}$ definitions.
%Setting $\sigma_0$ and $E_0$ to $\dot{\sigma_0} = 1 1/s$ and $E_0 = 10^4 V/m$ will result in fixed values of $Mn=70$, $Ca=80$ and $Re=5\times 10^{-4}$ in
 %the momentum equations. 

To demonstrate the robustness and effectiveness of the hydrodynamic portion of the numerical method and for the purpose of the verification, sample results are presented 
for a vesicle under linear shear flow in the absence of electric fields ($Mn=0$). Note that the characteristic time scale associated with for this case is $t_0=t_{\gamma}=1 s$.
Using the typical experimental values given in Sec. \ref{sec:2.5} the dimensionless parameters are found to be $Re= 0.001$ and $Ca= 10$. Unless otherwise stated, these parameters
along with a dimensionless shear rate of $\chi=1$ at the boundary are used for the hydrodynamic simulations as well as the results for combined effect of shear flow and the electric field.

Here two major modes of motions for the vesicle in shear flow are reported: tank-treading and tumbling.
The tank-treading motion happens if the viscosity ratio, $\eta$, is less than a critical value $\eta_c$, which depends on the reduced volume of the vesicle \cite{biben2003}. 
In the tank-treading regime the vesicle reaches an equilibrium angle and stays at that position. 
When the viscosity ratio is above the critical viscosity the behavior of the vesicle changes, and it starts to tumble end-over-end.
Studies have shown that the dynamics depend primarily on the viscosity ratio, $\eta$ and
the reduced volume, $v$, with no remarkable dependence on the shear rate strength, $\chi$ \cite{veerapaneni2009}.

\begin{figure}

		\begin{center}
			\subfigure[]{
				\includegraphics[width=0.45\textwidth]{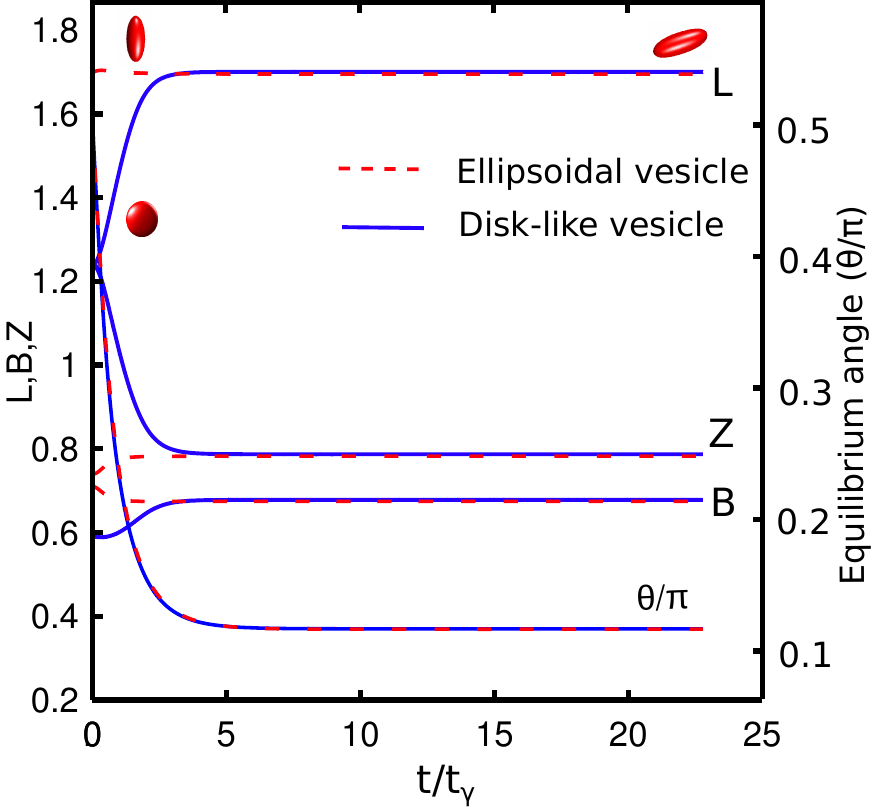}
				 \label{fig:TTLBZ}
			}
			\subfigure[]{
				\includegraphics[width=0.40\textwidth]{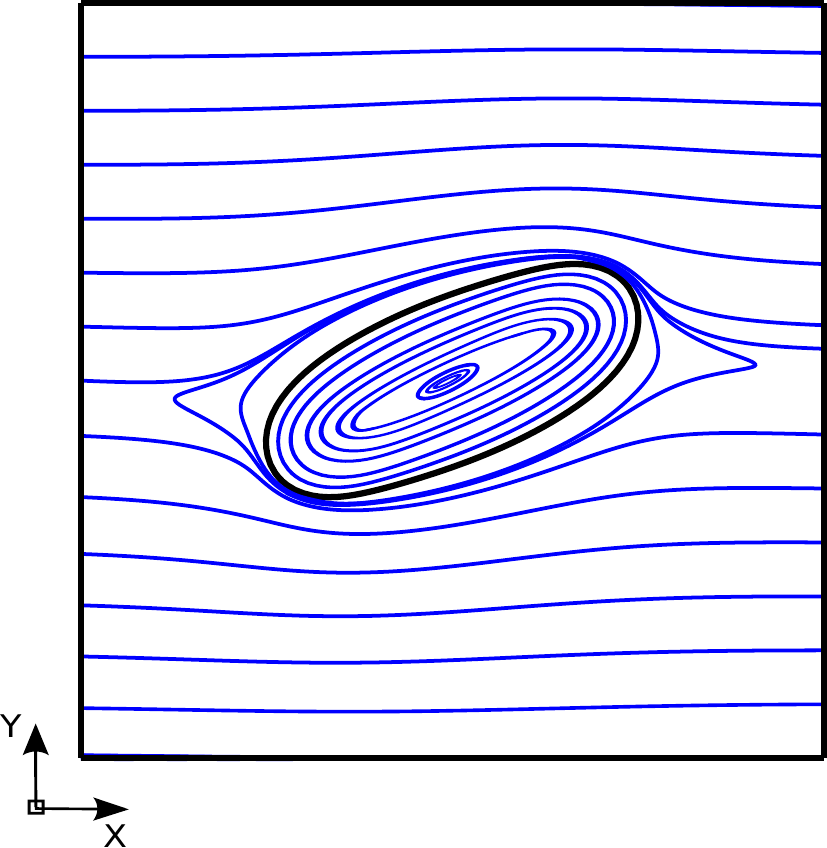}
				\label{fig:streamline}
			} 
		
		\end{center}
	\caption{Sample result of tank treading vesicle with different initial shapes.
	 The final equilibrium shape and angle are the same for both initially prolate and oblate vesicles.}
		\label{fig:TTanalysis} 
\end{figure}

Sample tank-treading results are illustrated in Fig. \ref{fig:TTLBZ} for two initial conditions: a disk-like vesicle and an ellipsoidal vesicle. Both initial conditions have 
the same reduced volume of $v=0.85$ and the dimensionless shear rate at the boundary is set to $\chi=1$.
The inclination angle with respect to the shear-flow axis, $\theta$, the semi-major axis $L$, 
semi-minor axis $B$, and the half axis length in the vorticity direction, $Z$,  are reported. These values are computed
through the use of the inertia matrix of the vesicle \cite{laadhari2014computing}. The eigenvalues of the interia matrix
correspond to the axis lengths, while the angle between the eigenvector associated with the largest eigenvalue and the $y-$axis 
is the inclination angle.

It is clear from Fig. \ref{fig:TTLBZ} that both initial conditions result in the same equilibrium shape and inclination angle.
Figure \ref{fig:streamline} shows the vortices  being formed in the interior of the vesicle at the equilibrium state. The streamlines demonstrate that the fluid velocity is tangent to the membrane which indicates
the tank-treading behavior.

\begin{figure}

		\begin{center}
			\subfigure[]{
				\includegraphics[width=0.45\textwidth]{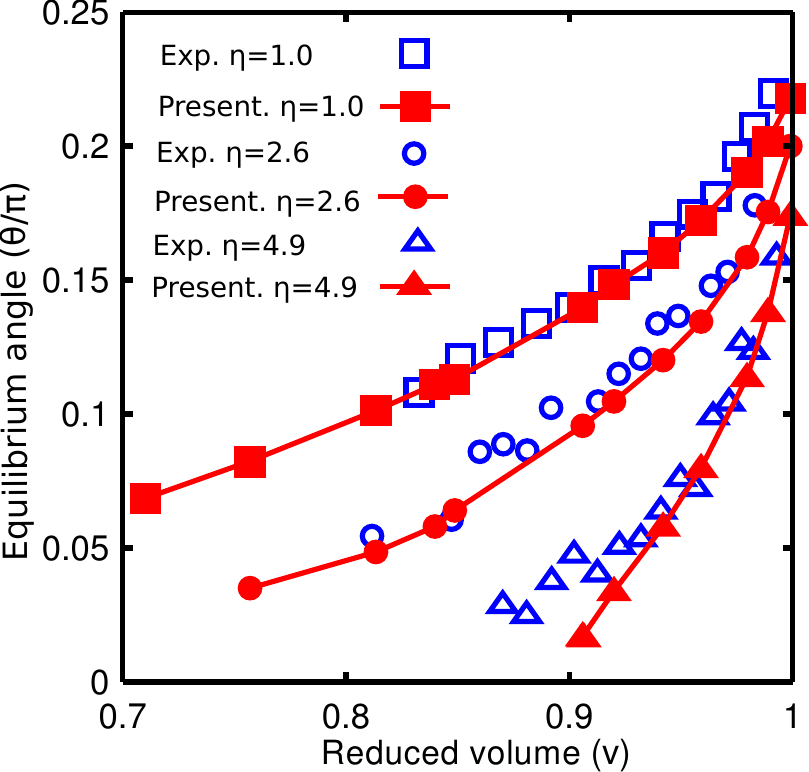}
				 \label{fig:TTexp}
			}
			\subfigure[]{
				\includegraphics[width=0.42\textwidth]{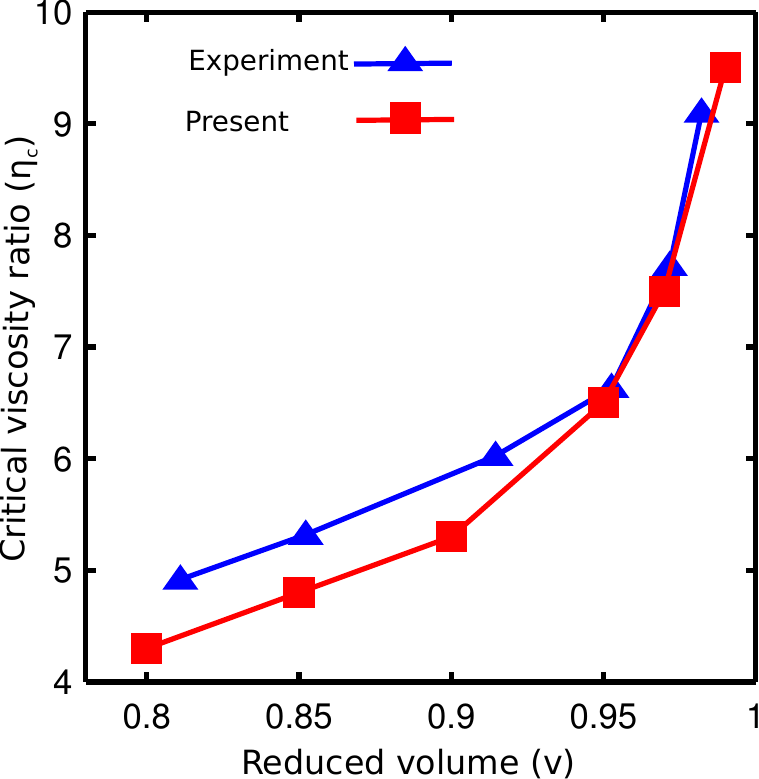}
				\label{fig:CritVisc}
			} 
		
		\end{center}
	\caption{Comparison of vesicle hydrodynamics to experiments. (a) The effects of reduced volume and viscosity ratio on the inclination angle of a 
	tank-treading vesicle. The results are compared against the experimental data from Ref. \cite{zabusky2011}. (b) 
	The critical viscosity ratio for vesicles with various reduced areas. The results are compared to the experimental data in Ref. \cite{kantsler2006}.}		
\end{figure}

The equilibrium inclination angle as a function of reduced volume and viscosity ratio is shown in Fig \ref{fig:TTexp}. 
A dimensionless shear rate of $\chi=1$ is applied at the boundary and results are compared to experimental data from Ref. \cite{zabusky2011}.
The angles from the numerical model are in good agreement with experimental measurements.
Small discrepancies in the case of $\eta=4.9$ could be related to the role of thermal fluctuations in the dynamics of vesicles with larger viscosity ratios \cite{zabusky2011}.
This effect is absent in the proposed model and further investigation is needed to fully understand this phenomena.

The transition between tank-treading and tumbling happens at the critical viscosity ratio $\eta_c$. As $v$ increases a larger $\eta_c$ is required to transition
from tank-treading to tumbling. This behavior is shown in Fig \ref{fig:CritVisc}.
The present method compares well to experimental data from \cite{kantsler2006}, particularly for larger reduced volumes.

%~ \begin{figure}[ht!]
	%~ \centering
	%~ \includegraphics[width=0.6\textwidth]{untitled}
	%~ \caption{Sample result }
	%~ \label{fig:streamline}
%~ \end{figure}
%~ 
%~ \begin{figure}[ht!]
	%~ \centering
	%~ \includegraphics[width=0.6\textwidth]{ProlateOblate_Ca50_Re01_Vrdot85}
	%~ \caption{Sample result of tank treading vesicle with different initial shapes.
	 %~ The final equilibrium shape and angle are the same for both initially prolate and oblate vesicles.}
	%~ \label{fig:TTLBZ}
%~ \end{figure}

%\begin{figure}[ht!]
	%\centering
	%\includegraphics[width=0.5\textwidth]{ExpTT_Ca10_Redot01}
	%\caption{The effects of reduced volume and viscosity ratio on the inclination angle of a 
	%tank-treading vesicle. The results are compared against the experimental data from Ref. \cite{zabusky2011}}.
	%\label{fig:TTexp}
%\end{figure}

%\begin{figure}[ht!]
	%\centering
	%\includegraphics[width=0.5\textwidth]{CriticalVisc}
	%\caption{The critical viscosity ratio for vesicles with various reduced areas and a shear rate of $\dot{\gamma}=0.125$. The results are compared to the experimental data in Ref. \cite{kantsler2006}}
	%\label{fig:CritVisc}
%\end{figure}

\subsection{Vesicle dynamics in strong DC electric field}
\label{sec:EHD}

The response of a vesicle to an external electric field will depend on a number of factors, including the membrane properties, the viscosity ratio, the electrical properties
of the fluids, and the applied electric field.
The shape of the vesicles is described by the deformation parameter:
\begin{align}
	D =& \frac{a_y}{a_x},	
\end{align}
where $a_y$ is the dimension of the vesicle in the direction parallel to the applied electric field (the $y$-direction) while $a_x$ is the
dimension of the vesicle in the direction perpendicular to the electric field, the $x$-direction. Note that these parameters are being calculated directly
using the interface location and differ from the semi-axis values defined in Sec. \ref{sec:HD}.

If the conductivity of the inner fluid is lower than the outer fluid, $\lambda<1$, and the electric field is strong enough such that $t_{ehd}<t_m$, 
an initially prolate vesicle will undergo a prolate-oblate-prolate (POP) transition \cite{salipante2014,schwalbe2011}. An oblate shape is given by $D<1$ while
a prolate shape has $D>1$.

\begin{figure}[ht!]
	\centering
	\includegraphics[width=0.75\textwidth]{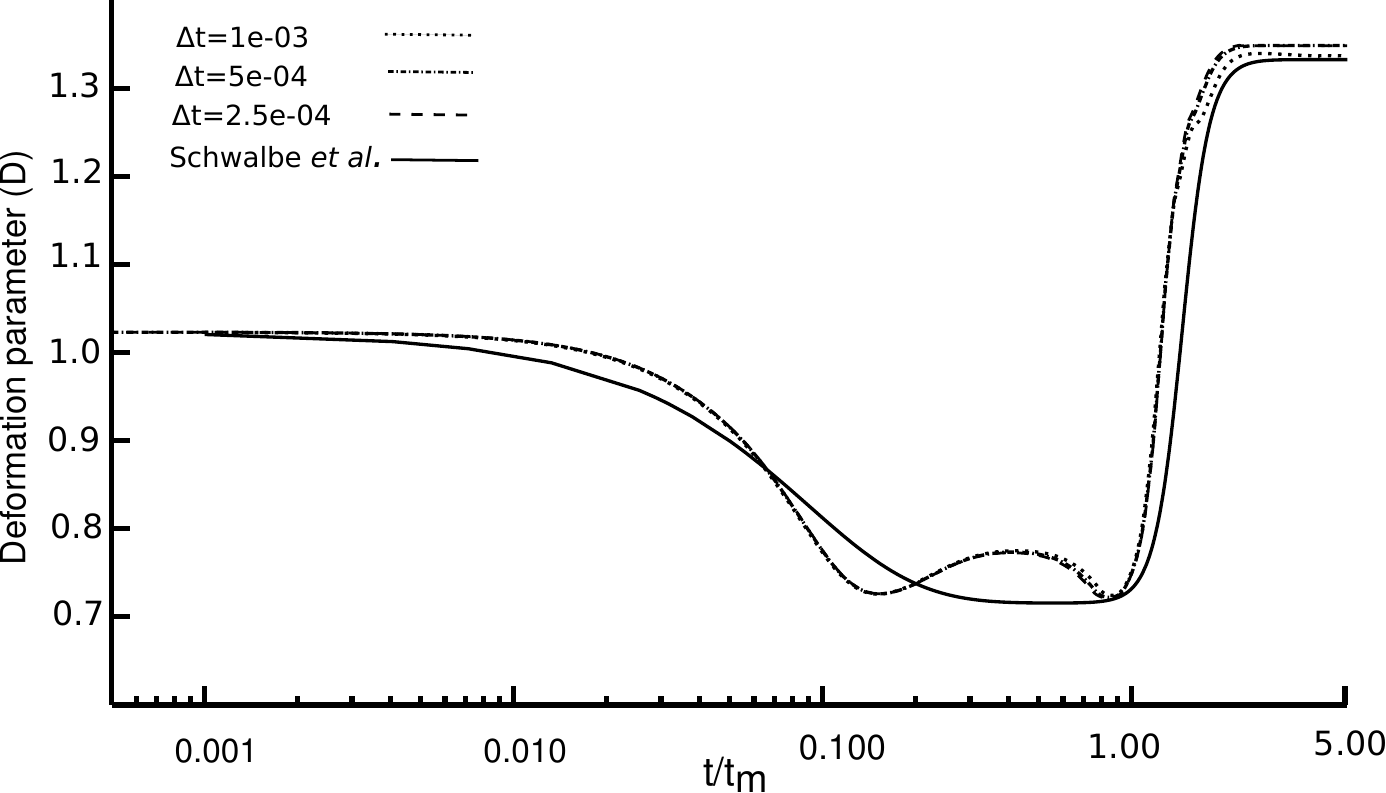}
	\caption{Comparison of the method against analytic solution of Schwalbe \textit{et. al.} \cite{schwalbe2011}. 
	Convergence test is done for three different time steps to check the temporal independence. The domain size is $[-4.5,4.5]$ with a grid spacing of $h=0.075$. The analysis
	demonstrates that $\Delta t=5\times10^{-4}$ is the maximum acceptable time-step. }
	\label{fig:Temporal}
\end{figure}

\begin{figure}[ht!]
	\centering
	\includegraphics[width=0.75\textwidth]{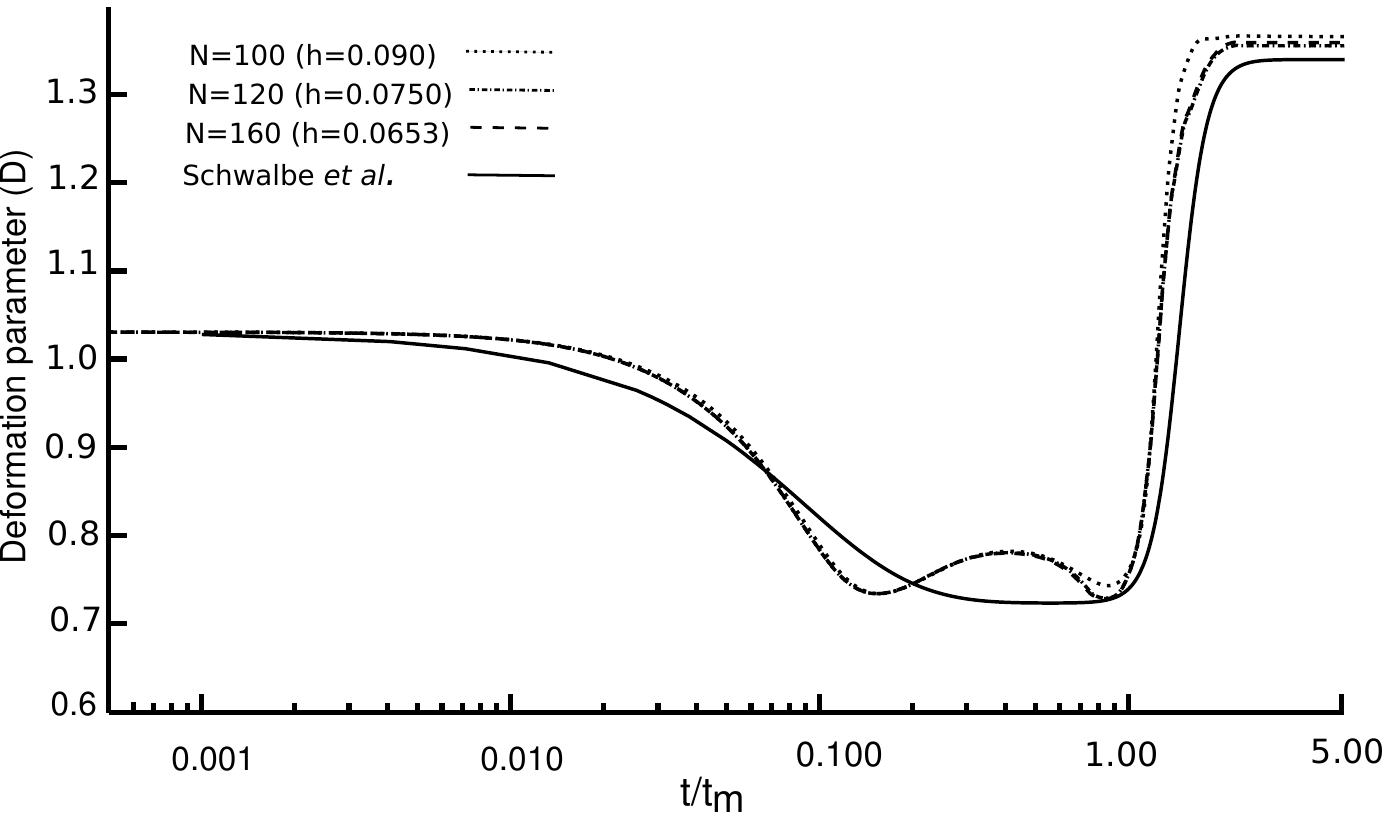}
	\caption{Comparison of the method against analytic solution of Schwalbe \textit{et. al.} \cite{schwalbe2011}. 
	Convergence test is done for three different grid spacing to check the grid independence. The domain size is $[-4.5,4.5]$ and the time step is set to  $\Delta t=5\times10^{-4}$.
	The investigation reveals that  that $h=0.075$ is the maximum acceptable grid spacing.}
	\label{fig:grid}
\end{figure}

\begin{figure}[ht!]
	\centering
	\includegraphics[width=0.75\textwidth]{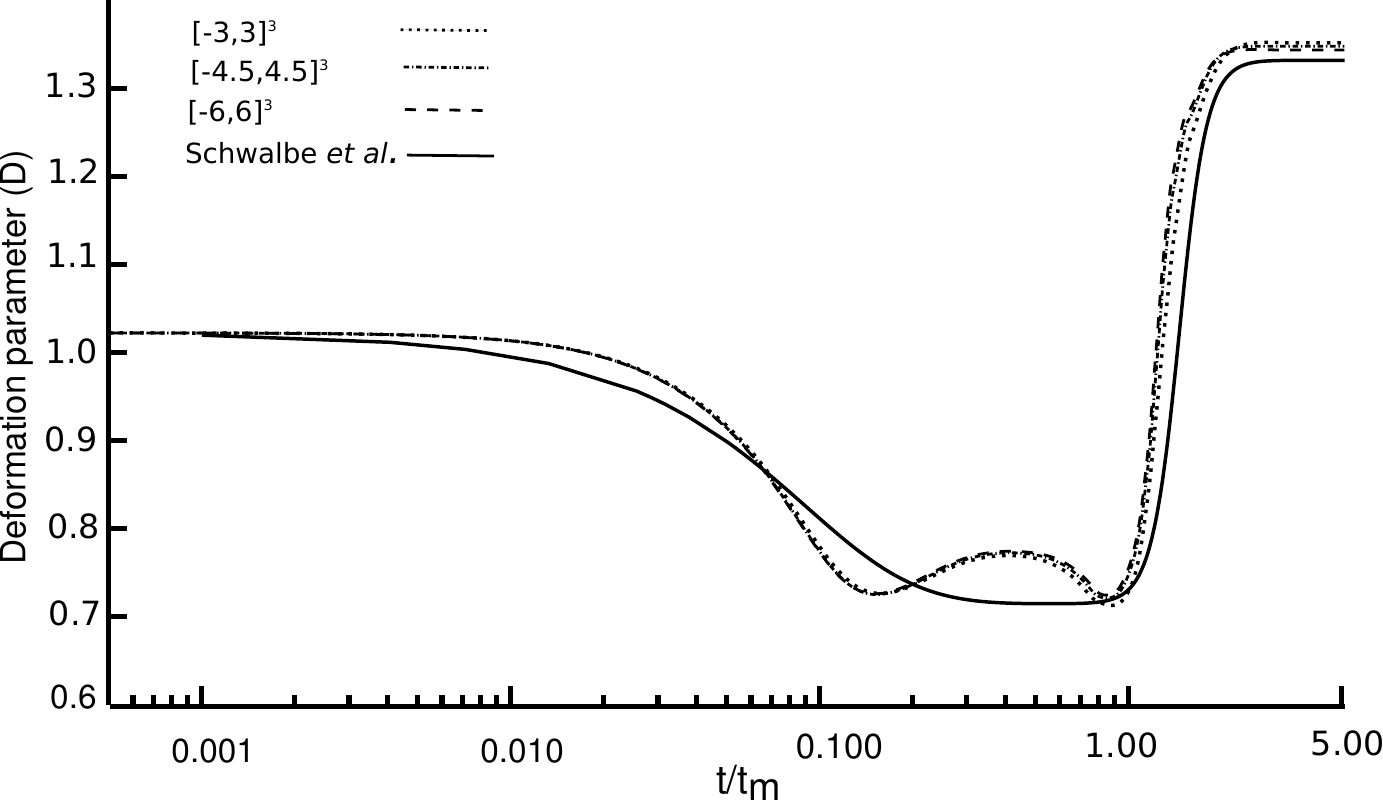}
	\caption{Comparison of the method against analytic solution of Schwalbe \textit{et. al.} \cite{schwalbe2011}. 
	Convergence test is done for three different domain sizes to check the domain confinement effect. The grid spacing is $h=0.075$ and the time step is set to  $\Delta t=5\times10^{-4}$.
	 Numerical results shows that $[-4.5,4.5]$ is the smallest acceptable domain size.}
	\label{fig:confinement}
\end{figure}

As verification of the method simulation results are compared to the analytic work of Schwalbe \textit{et. al.} \cite{schwalbe2011}.
The reduced volume of the vesicle is $v=0.98$. The initial shape is given by a second-order Spherical Harmonics parametric surface, see Ref. \cite{schwalbe2011} for details.
In Figs. \ref{fig:Temporal}-\ref{fig:confinement} the influence of the time step, grid size, and domain size is explored and results are compared to Ref. \cite{schwalbe2011}.
Note that the membrane charging time is chosen as the characteristic time in these simulations. Recall that using the typical experimental values of the physical parameters 
discussed earlier in Sec.\ref{sec:2.5} this results in $t_0=t_m=2.1\times10^{-3}$ s, $\hat{C}_m=0.095$, $\hat{G}_m=0$, $Re=0.19$, $Ca=3.8\times10^{4}$, $Mn=18$, $E_0=1$,  and $\chi=0$.
Also a matched dielectric ratio is used in all the electrohydrodynamic simulations in this paper.

\begin{figure}

		\begin{center}
			\subfigure[$t/t_m=0$]{
				\includegraphics[width=0.225\textwidth]{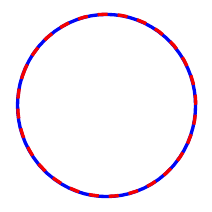}
				 \label{fig:SchwalbeXYcomp_0}
			}
			\subfigure[$t/t_m=0.06$]{
				\includegraphics[width=0.225\textwidth]{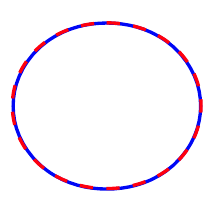}
				\label{fig:SchwalbeXYcomp_120}
			} 
			\subfigure [$t/t_m=0.14$]{
				\includegraphics[width=0.225\textwidth]{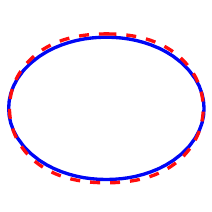}
				\label{fig:SchwalbeXYcomp_280}
			}
			\subfigure [$t/t_m=0.50$]{
				\includegraphics[width=0.225\textwidth]{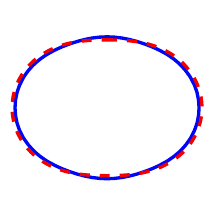}
				\label{fig:SchwalbeXYcomp_1000}
			}\\
			\vspace{0.25in}
			\subfigure [$t/t_m=0.90$]{
				\includegraphics[width=0.225\textwidth]{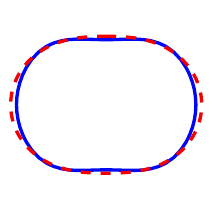}
				\label{fig:SchwalbeXYcomp_1800}
			}
			\subfigure [$t/t_m=1.20$]{
				\includegraphics[width=0.225\textwidth]{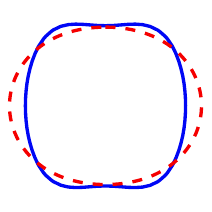}
				\label{fig:SchwalbeXYcomp_2600}
			}		
			\subfigure [$t/t_m=2.90$]{
				\includegraphics[width=0.225\textwidth]{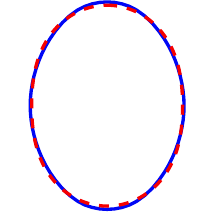}
				\label{fig:SchwalbeXYcomp_5800}
			}
			\subfigure [$t/t_m=5.00$]{
				\includegraphics[width=0.225\textwidth]{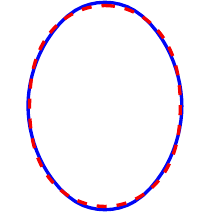}
				\label{fig:SchwalbeXYcomp_10000}
			}

		\end{center}
		\caption{ Contours of vesicle shape in the x-y plane over time. The solid blue line is the solution of the present method while the
		 dashed red line shows the solution of Schwalbe \textit{et. al.} \cite{schwalbe2011}. A clear difference is observed in the oblate-prolate
		 transition between the two solutions. While the analytic shape remains ellipsoidal during the evolution, the current method predicts a nearly cylindrical deformation
		 for the vesicle.}
		\label{fig:SchwalbeXYcomp} 
\end{figure}

Several points need to be made about the results. First, the use of a time step of size $\Delta t=5\times10^{-4}$, grid size of $N=120^3$ with a domain of $[-4.5,4.5]^3$ results
in a converged solution and is therefore used in the following simulations. Second, there is a discrepancy between the simulation results and the analytic results 
of Schwalbe \textit{et. al}. While the general results are similar, for example the approximate time at which the prolate-oblate transition occurs and the time
in the oblate shape, there are important differences. The transition from the initial prolate to oblate shape is delayed and is sharper for the current simulation
versus the analytic result. Also, the deformation parameter increases slightly during the oblate shape (at a time of 0.4$t_m$) before decreasing again. Finally,
the final equilibrium shape is slightly different between the two methods, as demonstrated by the difference in the final deformation parameter. 
This difference is most likely due to the use of a mode-2 spherical harmonic description of the interface in the analytic work. The current simulation has access to a wider range
of vesicle shapes than the analytic work and thus additional deformation modes will be accessible. This can be seen by comparing the cross-section of the vesicle 
in the $x-y$ plane for the two results, Fig. \ref{fig:SchwalbeXYcomp}. The analytic works always remains ellipsoidal in shape while 
the numerical simulation developed here demonstrates the availability of additional shapes. The other possible explanation for the discrepancy between the two results would be related to
the use of Stokes assumption in the analytic solution. The present method uses a full Navier-Stokes solver and therefore possible inertia effects are taken into account.

\begin{figure}

		\begin{center}
			\subfigure[$t/t_m=0$]{
				\includegraphics[width=0.225\textwidth]{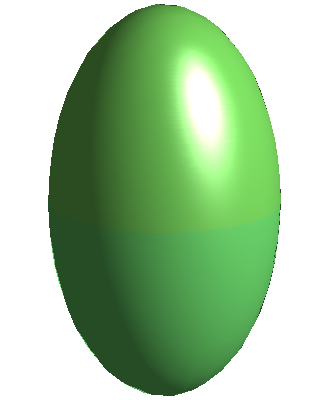}
			}
			\subfigure[$t/t_m=0.25$]{
				\includegraphics[width=0.225\textwidth]{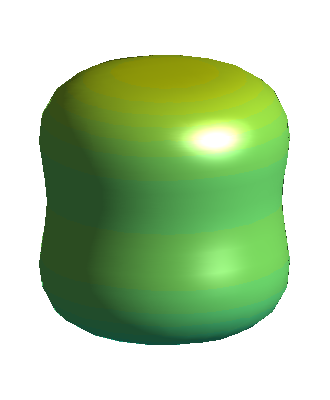}
			} 
			\subfigure [$t/t_m=0.35$]{
				\includegraphics[width=0.225\textwidth]{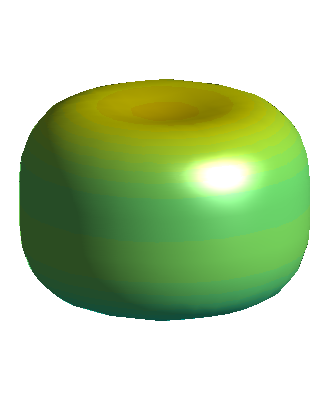}
			}
			\subfigure [$t/t_m=0.65$]{
				\includegraphics[width=0.225\textwidth]{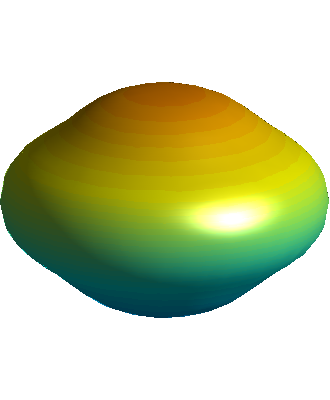}
			}\\
			\vspace{0.25in}
			\subfigure [$t/t_m=1.00$]{
				\includegraphics[width=0.225\textwidth]{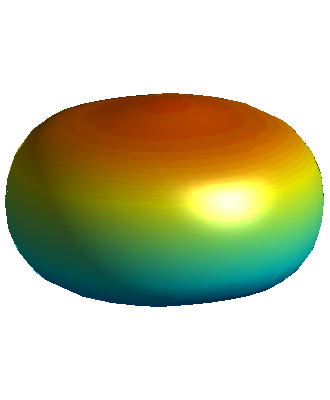}
			}
			\subfigure [$t/t_m=1.40$]{
				\includegraphics[width=0.225\textwidth]{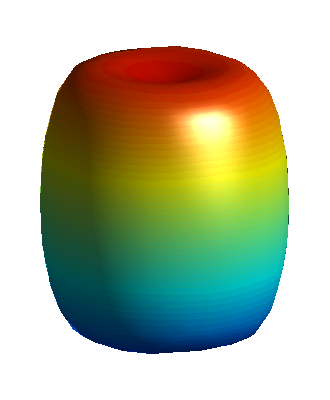}
			}		
			\subfigure [$t/t_m=2.50$]{
				\includegraphics[width=0.225\textwidth]{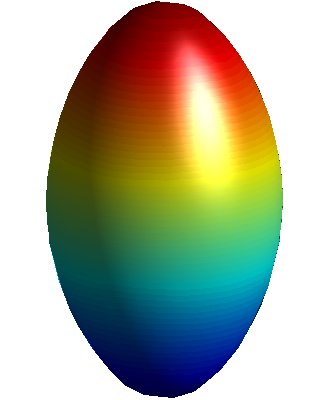}
			}
			\subfigure [$t/t_m=5.00$]{
				\includegraphics[width=0.225\textwidth]{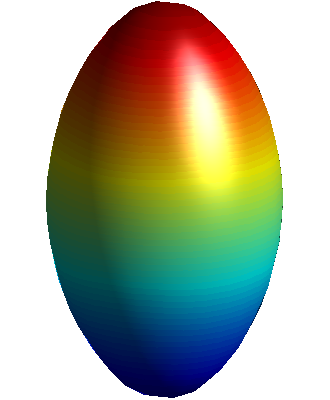}
			}

		\end{center}
		\caption{POP transition for the vesicle exposed to strong uniform DC electric field. The vesicle has a reduced volume of $v=0.93$, membrane capacitance of $\hat{C}_m=0.095$
		and conductance of $\hat{G}_m=0$. The fluid conductivity ratio is $\lambda=0.1$ with matched viscosity and dielectric ratios. The dimensionless parameters
		are given as $Mn=35$, $Re=0.19$ and $Ca=38000$. The colors (online version) indicate the trans-membrane potential, with blue (bottom of vesicle) indicating 
		a membrane potential of $V_m=-1.6$ and red (top of vesicle) indicating a membrane potential of $V_m=-1.6$.}
		\label{fig:3d_POP} 
\end{figure}

To demonstrate the axisymmetric POP behavior an initially prolate vesicle with reduced volume of $v=0.93$ and matched viscosity ($\eta=1$) is placed in a strong electric DC field.
All the dimensionless parameters except for the Mn are the same as before.
To ensure the full POP transition for the more deflected vesicle the Mason number is set to $Mn=35$.
The electrohydrodynamic time for this situation is computed as
$t_{ehd}=1.2\times10^{-4}s <t_m$ and therefore the POP transition is expected.
Three-dimensional results are shown in Fig. \ref{fig:3d_POP} at various time during the evolution. The trans-membrane potential on the interface
is also shown, which demonstrates the charging process.

The results clearly demonstrate the transition from a prolate shape to an oblate shape, and then back up to prolate. The initial prolate-oblate
transition occurs at about the membrane charging time. This type of behavior has been previously predicted numerically \cite{schwalbe2011,mcconnell2013} and verified 
experimentally \cite{salipante2014}. The cylindrical deformations observed in two-dimensional vesicle electrohydrodynamic simulations is also observed \cite{mcconnell2013}.
It is worth noting that the dimple at the center of the vesicle seen at times 0.35 $t_m$ and 1.40 $t_m$ are more pronounced than in the two-dimensional simulations.
This could be due to the different parameters used here and the fact that three-dimensional vesicles are typically more flexible due to the additional curvatures and 
deformation modes.

\begin{figure}[ht!]
	\centering
	\includegraphics[width=0.45\textwidth]{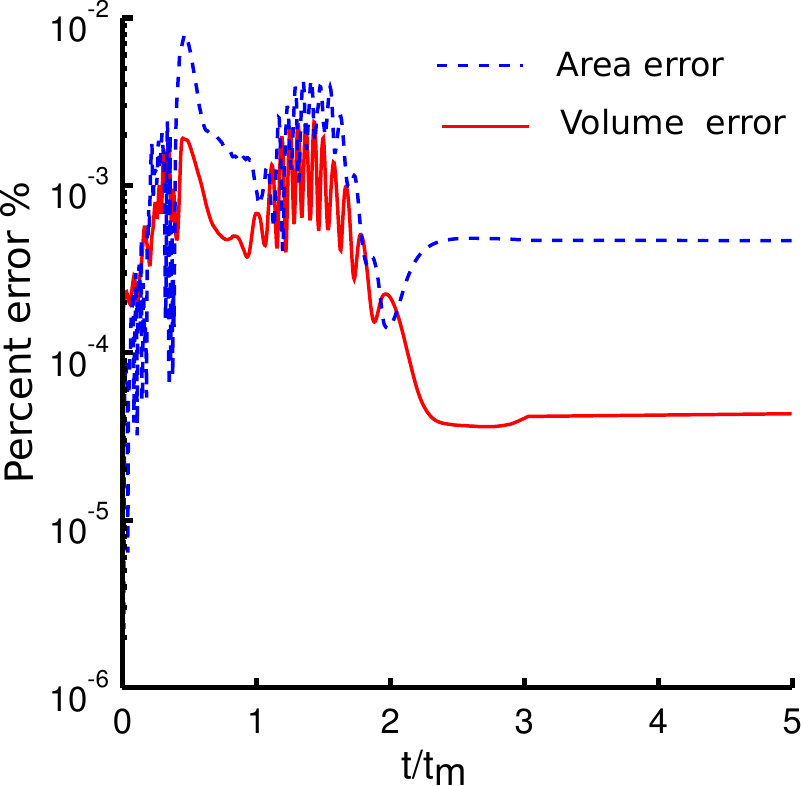}
	\caption{Conservation of the area and volume of the vesicle during the POP transition given in Fig. \ref{fig:3d_POP}. 
			 The percent errors never exceed $0.01\%$ over time.}
	\label{fig:AreaVolError}
\end{figure}

The area and volume were also tracked during this symmetric POP simulation and the errors, as a percentage of the initial area and volume, is given in Fig. \ref{fig:AreaVolError}. 
As can be seen both the area and volume are conserved extremely well.

\begin{figure}[ht!]
	\centering
	\includegraphics[width=0.9\textwidth]{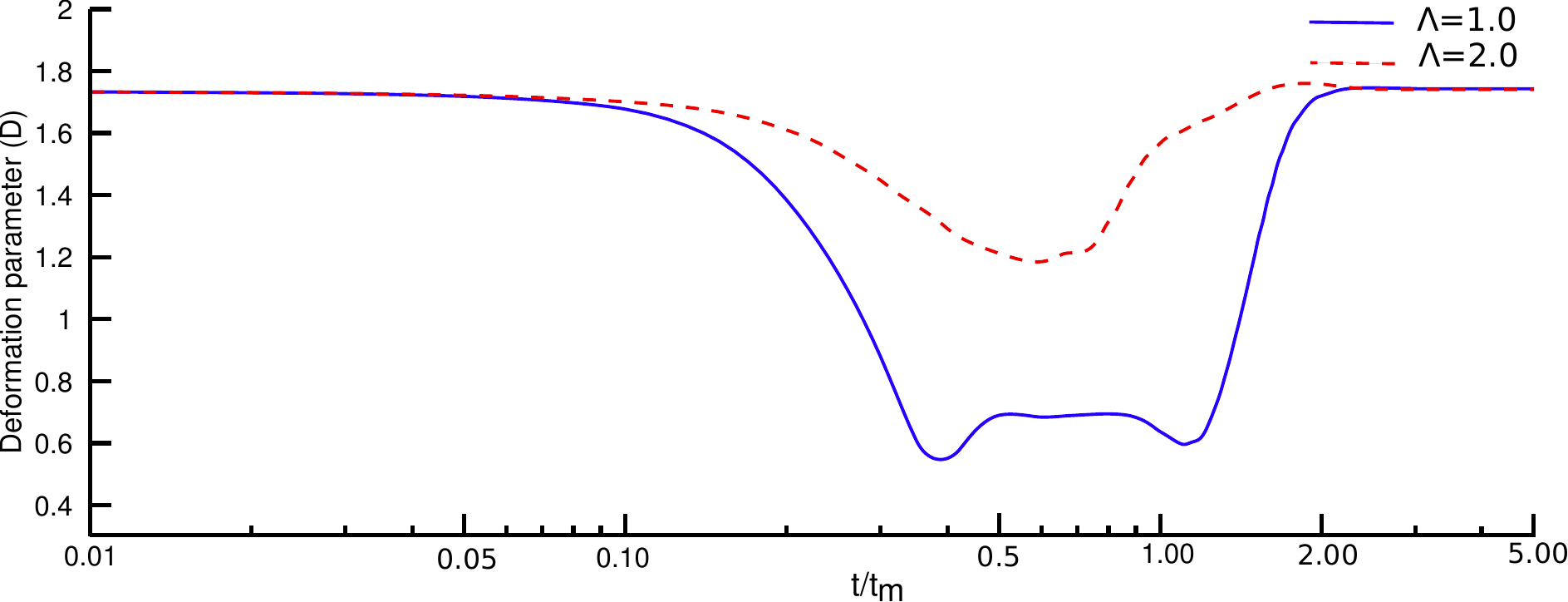}
	\caption{The effect of viscosity ratio on the dynamics of the vesicle. The solid line shows the deformation parameter for the vesicle with the information given in Fig. \ref{fig:3d_POP}. 
			This is compared against a vesicle with 
	        the same conditions but a viscosity ratio of $\eta=2.0$. While a full POP transition is observed for the case with matched viscosity ratio, the vesicle 
	        with larger $\eta$ never evolves into the oblate shape. }
	\label{fig:POPVisc} 
\end{figure}

The methods developed here allow for differing viscosity between the inner and outer fluids. As has been demonstrated previously an increase in the viscosity 
ratio can dramatically change vesicle dynamics in the absence of electric fields. Here the result seen in Fig. \ref{fig:3d_POP} is compared to a vesicle with the same parameters
except that the viscosity ratio is increased to $\eta=2$. The resulting deformation parameter evolution is presented in Fig. \ref{fig:POPVisc}. The increase 
in the viscosity ratio results in a vesicle which no longer undergoes a full prolate-oblate-prolate transition. Instead, the higher viscosity vesicle
has a slight flattening, but never reaches an oblate shape. In both cases the final equilibrium shape is the same. 
 
\subsection{Vesicle dynamics in combination of shear flow and DC field}
\label{sec:4.3}
When both a shear flow and DC electric field are applied the forces will compete to determine the dynamics of the vesicle. Previous works have demonstrated 
that electric fields damp the tank-treading and tumbling motion of vesicles in shear flow \cite{schwalbe2011, mcconnell2013}. 
If the electric field and shear flow directions are perpendicular the vesicle will have forces which act in perpendicular directions, leading to a competition 
between the two forces.

\begin{figure}[ht!]
	\centering
	\includegraphics[width=0.5\textwidth]{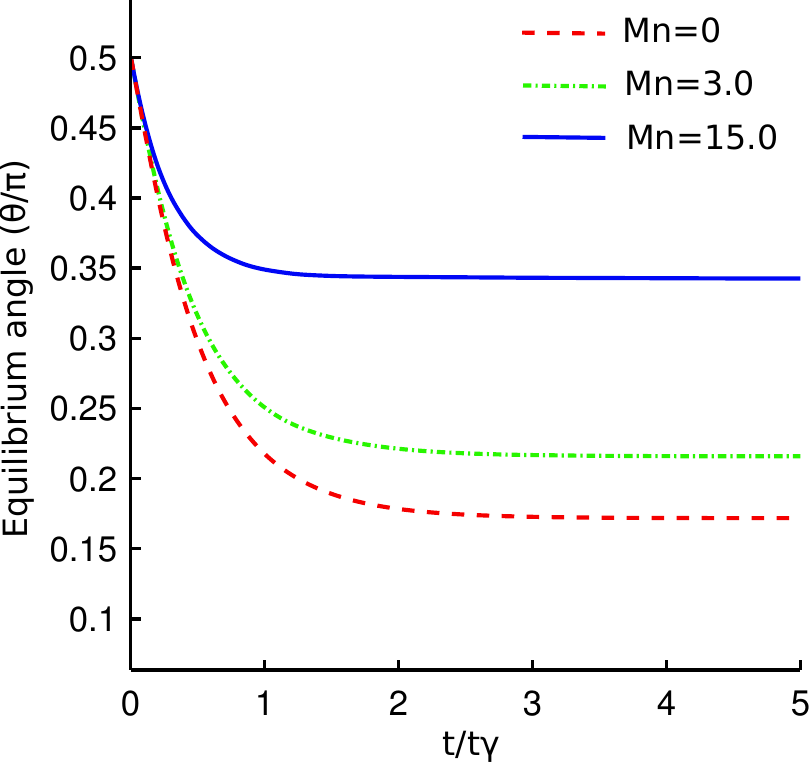}
	\caption{The comparison of inclination angle between standard tank-treading vesicle vs.
	vesicle in the presence of combined shear flow and weak electric fields. All the three cases
	use a matched viscosity ($\eta=1$) and a reduced volume of $v=0.93$. The rest of simulation parameters are $\hat{C}m=2\times10^{-4}$, $Re=0.001$, $Ca=10$ and $\chi=1$. 
	The membrane of vesicles with $Mn \neq 0$ 
	is initially charged. As the strength of the electric field increases the equilibrium vesicle angle increases.}
	\label{fig:TTefieldAngle}
\end{figure}

\begin{figure}[ht!]
        \begin{center}
            \subfigure[$t/t_{\dot{\gamma}}=0.0$] {
                  \includegraphics[width=0.225\textwidth]{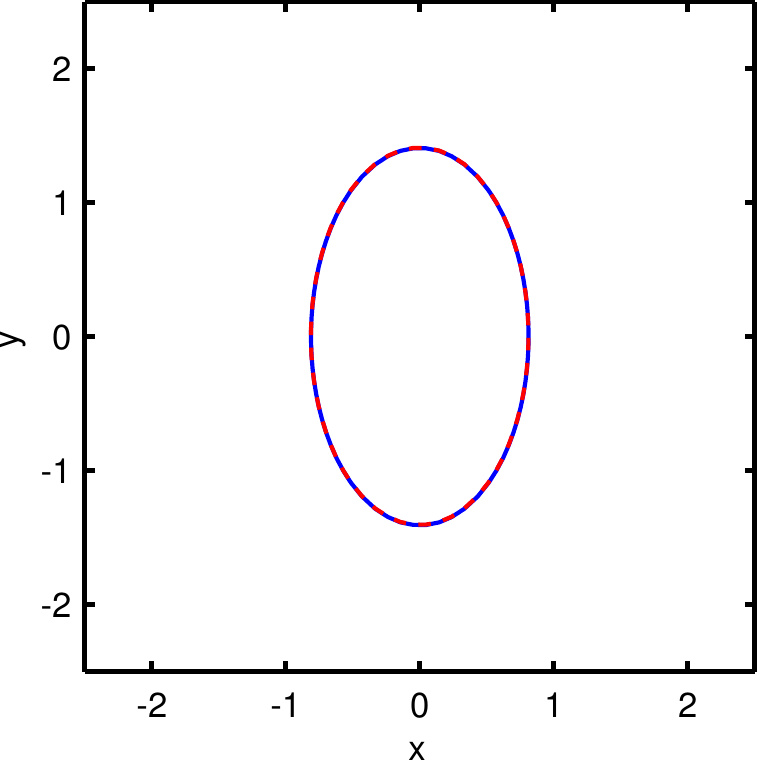}
                \label{fig:TTefield00}
            }             
            \subfigure[$t/t_{\dot{\gamma}}=0.5$]{
                  \includegraphics[width=0.225\textwidth]{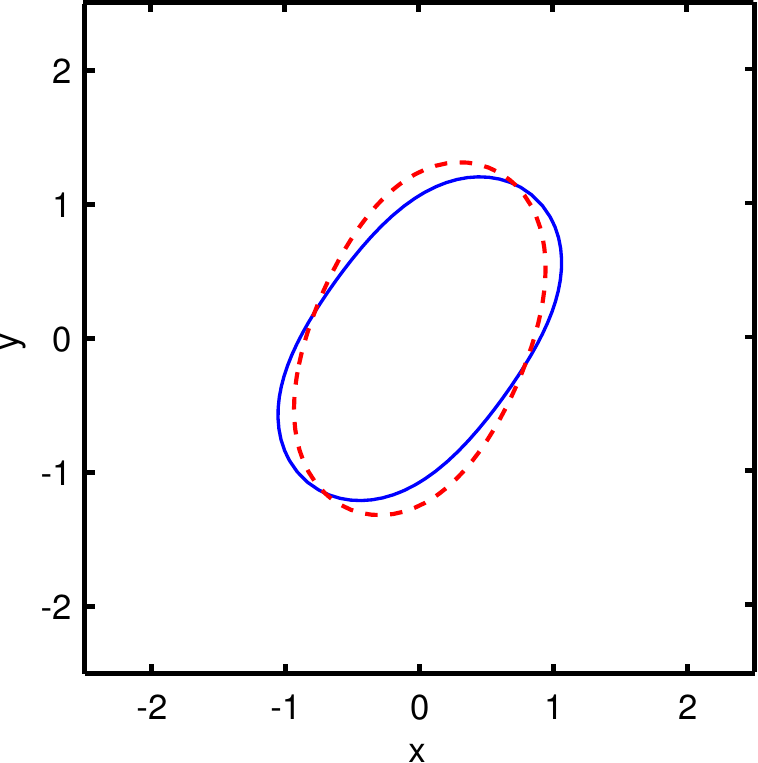}
                \label{fig:TTefield05}
            }
            \subfigure[$t/t_{\dot{\gamma}}=1.0$] {
                  \includegraphics[width=0.225\textwidth]{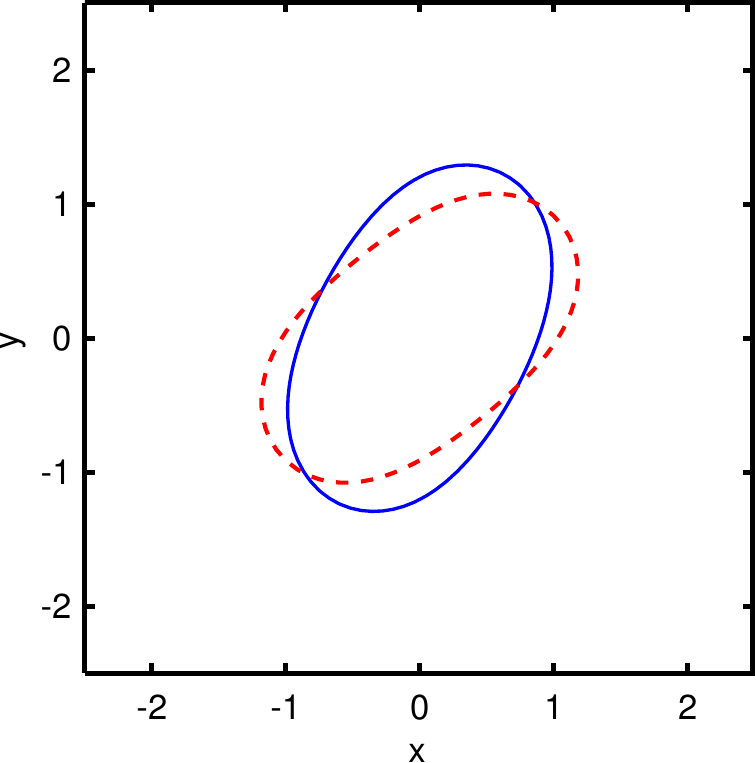}
                \label{fig:TTefield10}
			}
			\subfigure[$t/t_{\dot{\gamma}}=5.0$] {
                  \includegraphics[width=0.225\textwidth]{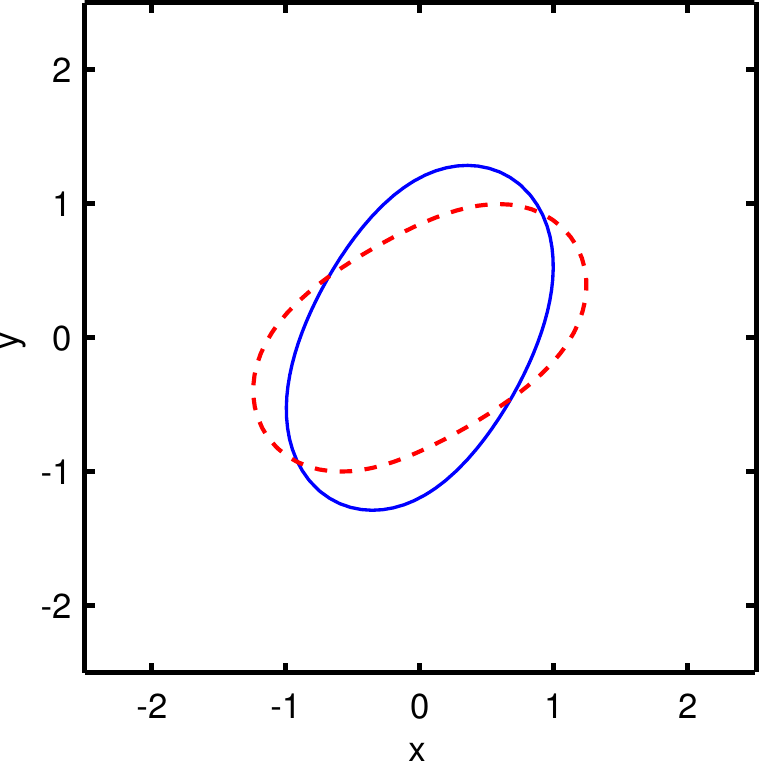}
                \label{fig:TTefield50}
			}
            \caption{The x-y cross section of the vesicle with initially charged membrane,$v=0.93$ and $\eta=1$ under combined effect of shear flow and weak DC field 
            with the strength of $Mn=15$ (the solid line; blue in the online version). 
            The dynamics is compared against a tank-treading vesicle with no
            external field, $Mn=0$ (dashed line; red in the online version). The rest of the parameters are $\hat{C}m=2\times10^{-4}$, $Re=0.001$, $Ca=10$ and $\chi=1$. 
            The presence of the electric field affects the inclination angle and causes the vesicle to reach
            the equilibrium condition at a smaller inclination angle. This behavior is due to the vertical alignment of the electric field in the y direction.
             The electric field force induces a resistance against any angular movement about the initial vertical position of the vesicle.}    
            \label{fig:TT_Efield}
        \end{center}
        
    \end{figure}

To demonstrate this consider a vesicle with matched viscosity, $\eta=1$ and a reduced volume of $v=0.93$ in a normalized shear flow of strength $\chi=1$. The characteristic time
is taken to be the shear flow timescale,$t_0=t_{\dot{\gamma}}=1$ s, and thus using standard membrane parameters the dimensionless coefficients 
become $\hat{C}m=2\times10^{-4}$, $\hat{G}_m=0$, $Re=0.001$, and $Ca=10$. The behavior of this vesicle under the influence of three
electric field strengths, $Mn=0$, $Mn=3$ and $Mn=15$ are shown in Fig. \ref{fig:TTefieldAngle}. Note that due to the membrane charging time, $t_m\approx 2.1\times10^{-3}$ s,
being much faster than the characteristic time it is assumed that the vesicle membrane begins fully charged. During each time step the 
trans-membrane potential is updated using a sub-step iteration until a pseudo-steady state is reached. As expected the application of an electric field 
results in the equilibrium inclination angle increasing. A comparison in the $x-y$ plane for the vesicles with $Mn=0$ and $Mn=15$ 
are seen in Fig. \ref{fig:TT_Efield}, which clearly demonstrates the higher inclination angle.
      
\label{sec:Shear_Efield}
 
Next consider the application of an electric field to a vesicle in the tumbling regime, $\eta=10>\eta_c$. The other parameters are the same
as the tank-treading case: $v=0.93$, $\chi=1$, $\hat{C}m=2\times10^{-4}$, $\hat{G}_m=0$, $Re=0.001$, and $Ca=10$. As in the tank-treading 
case the trans-membrane potential is iterated until a pseudo-steady state is reached every time step. 

\begin{figure}[ht!]
	\centering
	\includegraphics[width=0.5\textwidth]{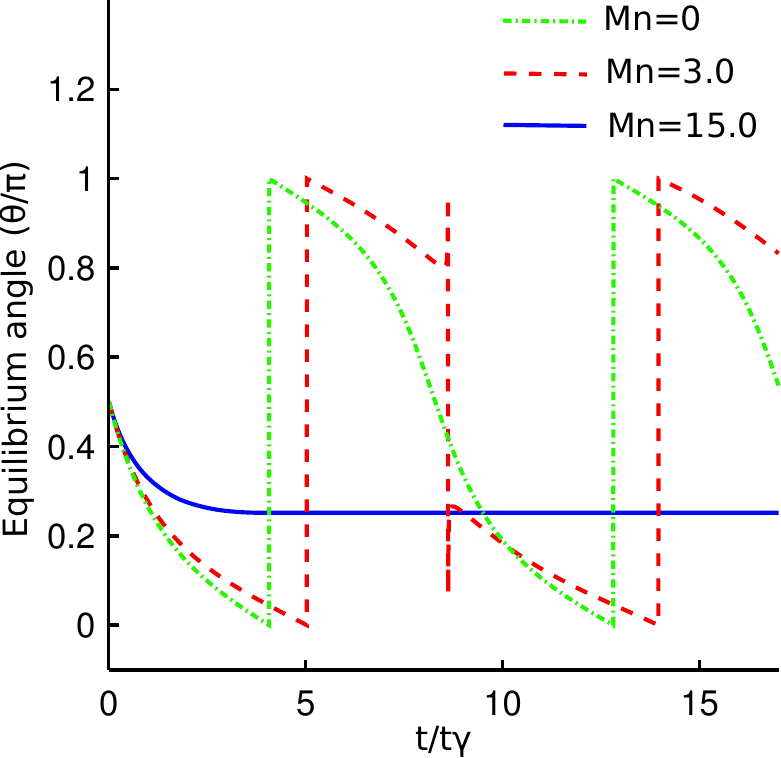}
	\caption{The comparison of dynamics between the standard tumbling vesicle vs.
	vesicle in the presence of combined shear flow and weak electric fields. All the three cases
	use a viscosity ratio of $\eta=10$ and a reduced volume of $v=0.93$. Other simulation parameters are $\hat{C}m=2\times10^{-4}$, $Re=0.001$, $Ca=10$ and $\chi=1$. 
	The membranes of vesicles with $Mn \neq 0$ 
	are initially charged. Compared to the standard tumbling case, the simulation with $Mn=3$ shows a lagged tumbling behavior and experiences a lot more 
	topological changes during the transition as it will be illustrated below. An interesting observation has been made for the vesicle with $Mn=15$. 
	In this situation, the vesicle undergoes a tank-treading motion and stays
	at that position permanently. }
	\label{fig:TBefieldAngle}
\end{figure}

\begin{figure}[ht!]
	\begin{center}
		\subfigure[$t/t_{\dot{\gamma}}=0.0$]{
			  \includegraphics[width=0.15\textwidth]{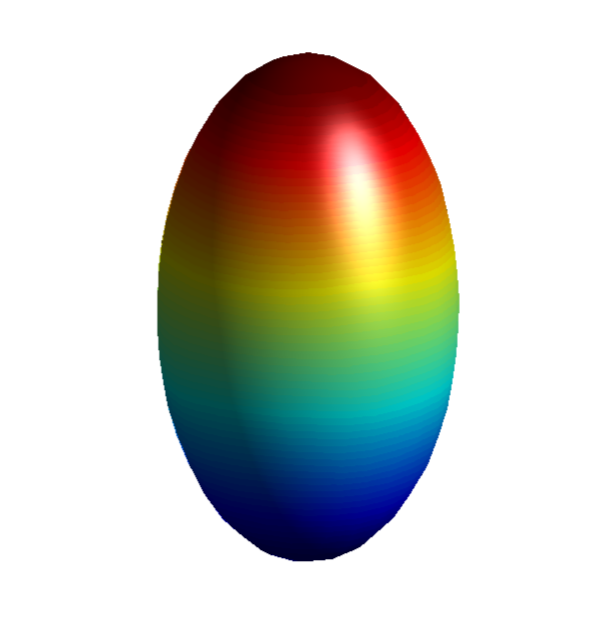}
			\label{fig:TBefield00}
		}             
		\subfigure[$t/t_{\dot{\gamma}}=2.5$]{
			  \includegraphics[width=0.15\textwidth]{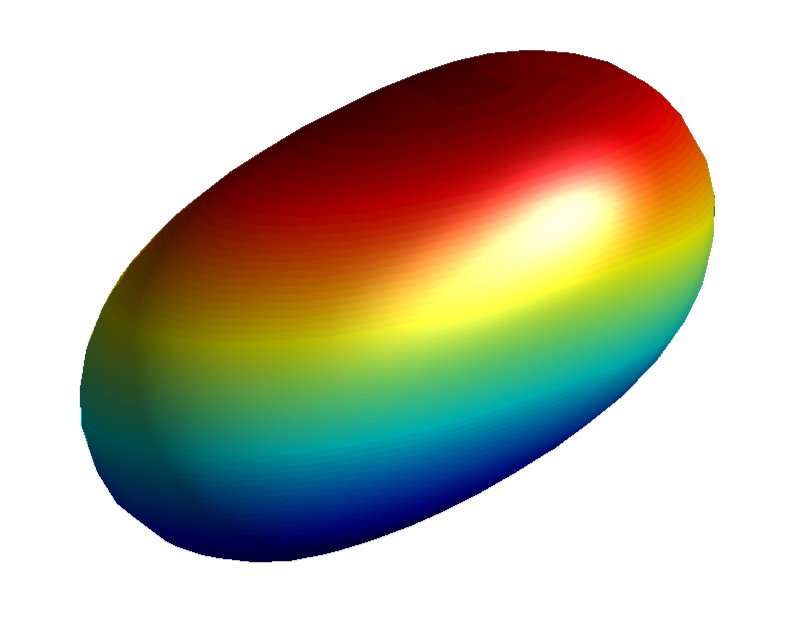}
			\label{fig:TBefield025}
		}
		\subfigure[$t/t_{\dot{\gamma}}=3.5$]{
			  \includegraphics[width=0.15\textwidth]{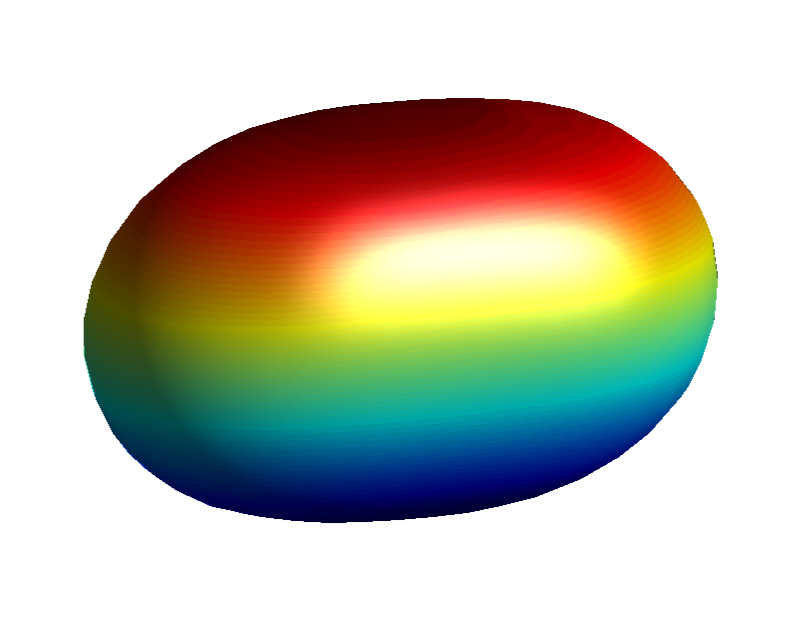}
			\label{fig:TBefield050}
		}
		\subfigure[$t/t_{\dot{\gamma}}=6.5$]{
			  \includegraphics[width=0.15\textwidth]{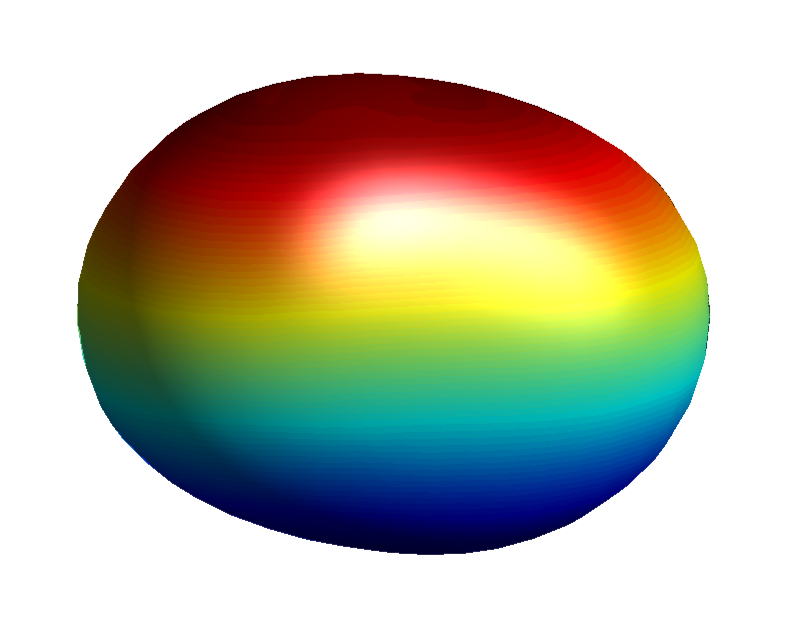}
			\label{fig:TBefield065}
		}
		\subfigure[$t/t_{\dot{\gamma}}=7.5$]{
			  \includegraphics[width=0.15\textwidth]{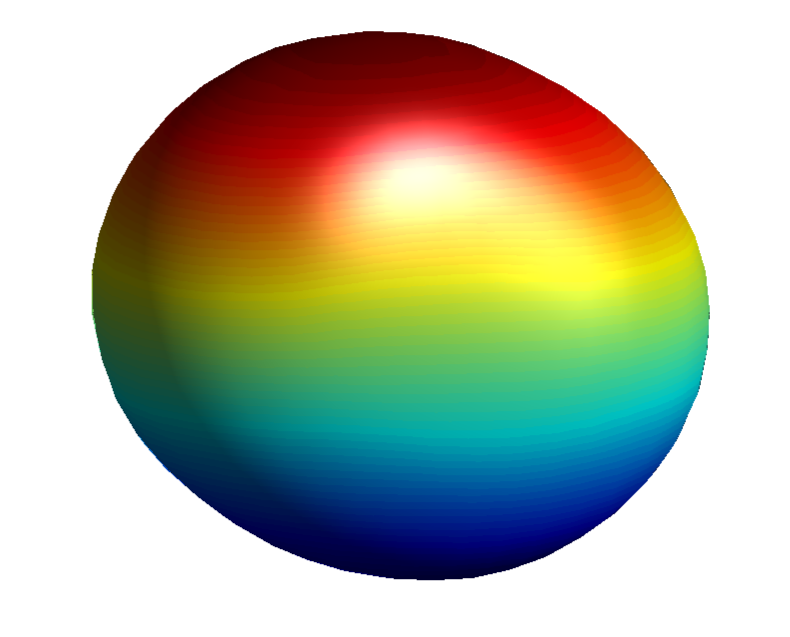}
			\label{fig:TBefield075}
		}
		\subfigure[$t/t_{\dot{\gamma}}=8.5$]{
			  \includegraphics[width=0.15\textwidth]{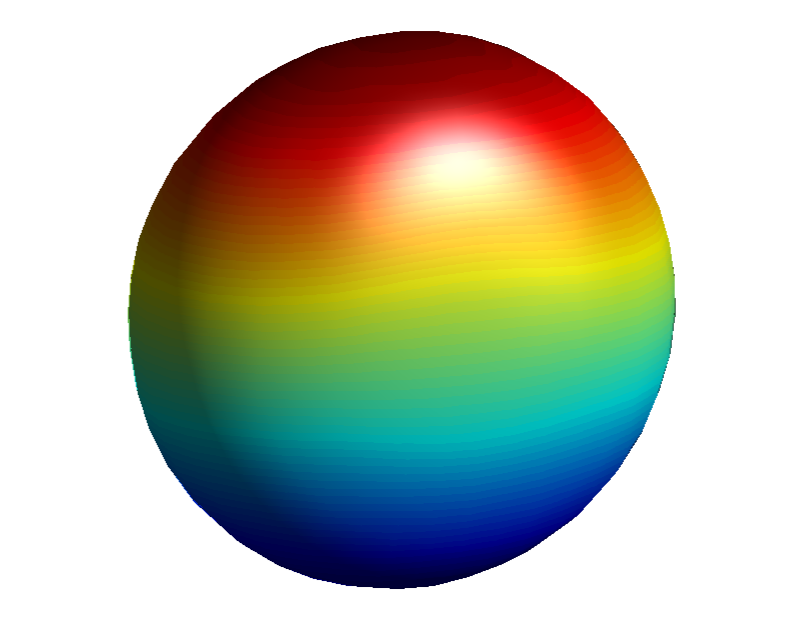}
			\label{fig:TBefield085}
		}             
		\subfigure[$t/t_{\dot{\gamma}}=10.0$]{
			  \includegraphics[width=0.15\textwidth]{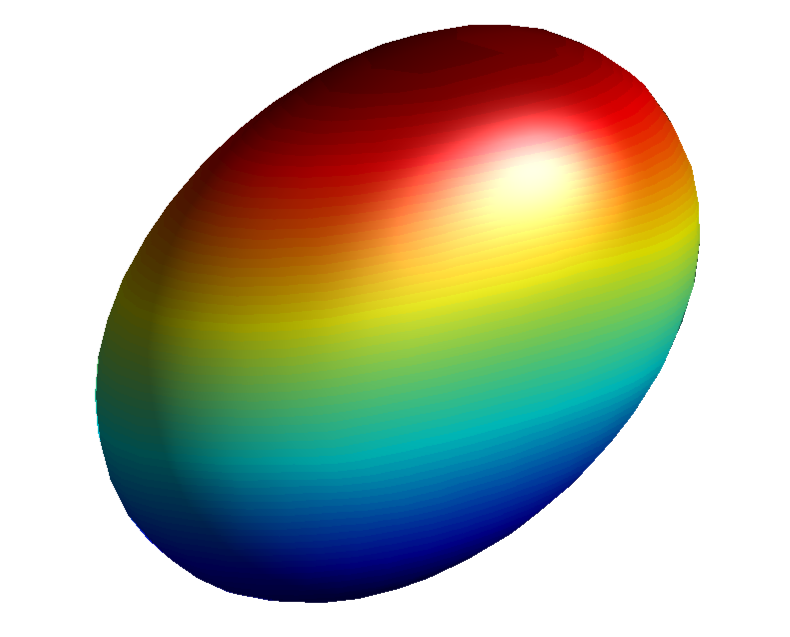}
			\label{fig:XYTBefield0100}
		}
		\subfigure[$t/t_{\dot{\gamma}}=12.5$]{
			  \includegraphics[width=0.15\textwidth]{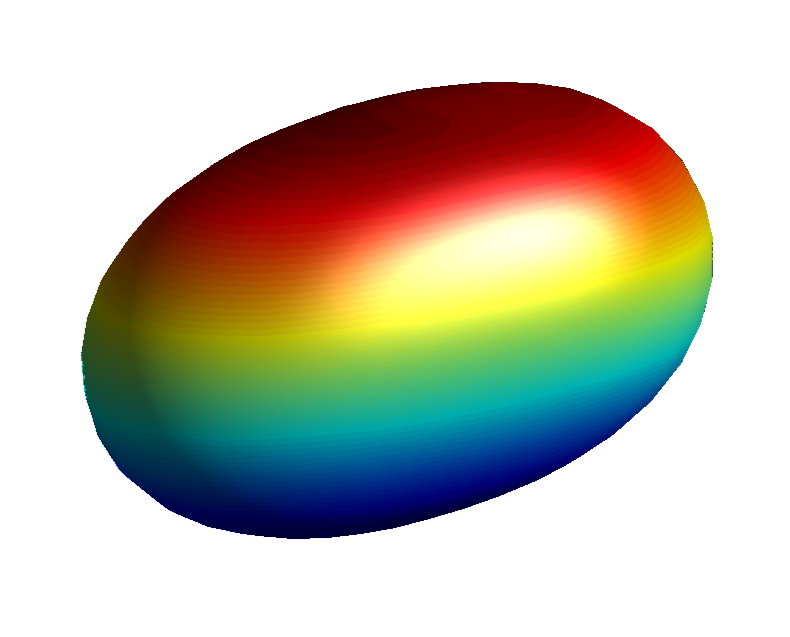}
			\label{fig:XYTBefield0125}
		}
		\subfigure[$t/t_{\dot{\gamma}}=15.0$]{
			  \includegraphics[width=0.15\textwidth]{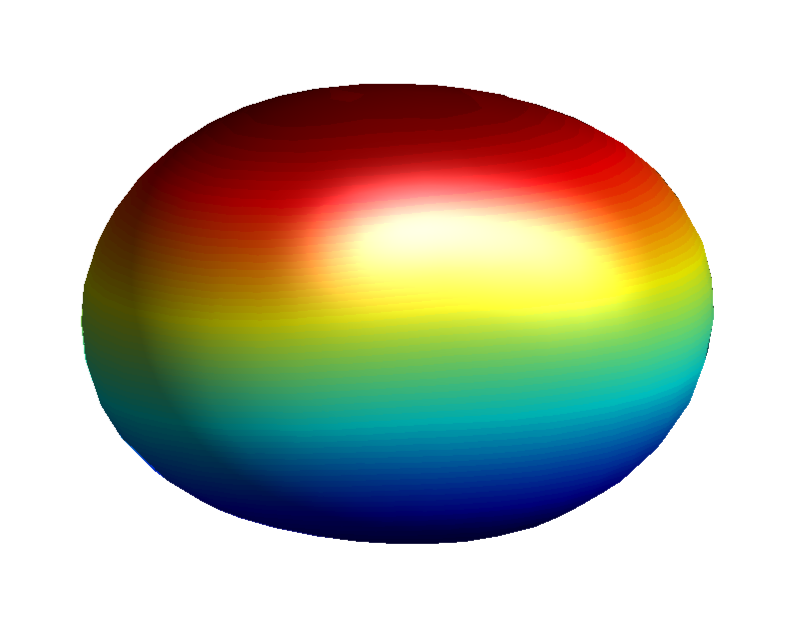}
			\label{fig:XYTBefield0150}
		}
		\subfigure[$t/t_{\dot{\gamma}}=17.0$]{
			  \includegraphics[width=0.15\textwidth]{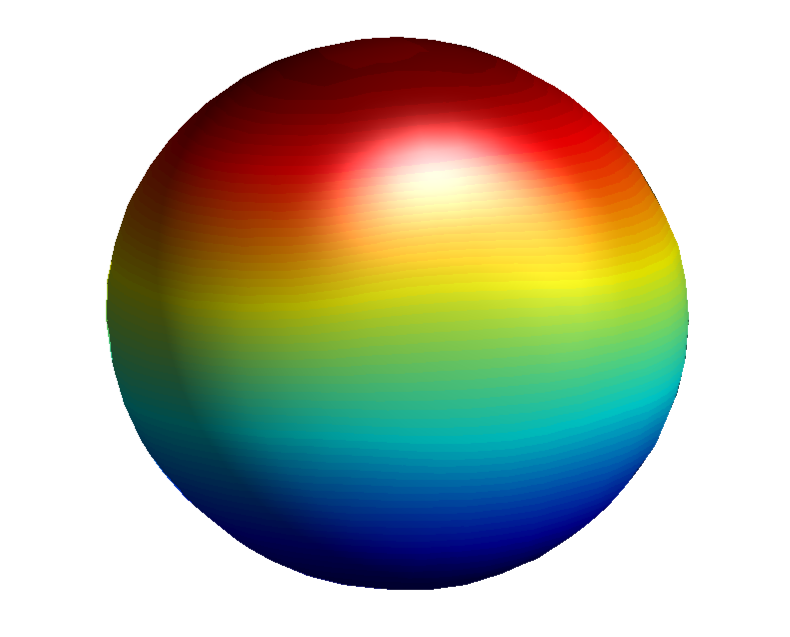}
			\label{fig:XYTBefield0170}
		}
		\caption{Simulation results on the dynamics of the vesicle with initially charged membrane, $v=0.93$ and $\eta=10$ under combined influence of shear flow and weak DC field.
		These results correspond to the case of $Mn=3$ in Fig. \ref{fig:XYTB_Efield}. The presence of the electric field influences the normal flipping motion of the vesicle.
		The topological changes between time $t=6.5$ to $t=10.0$ show that the vesicle first retracts at the poles prior to getting to the vertical direction and reforms back into
		the ellipsoidal shape passed the vertical position.
		}    

		\label{fig:TB_Efield}
	\end{center}
	
\end{figure}

\begin{figure}[ht!]
	\begin{center}
		\subfigure[$t/t_{\dot{\gamma}}=0.0$]{
			  \includegraphics[width=0.15\textwidth]{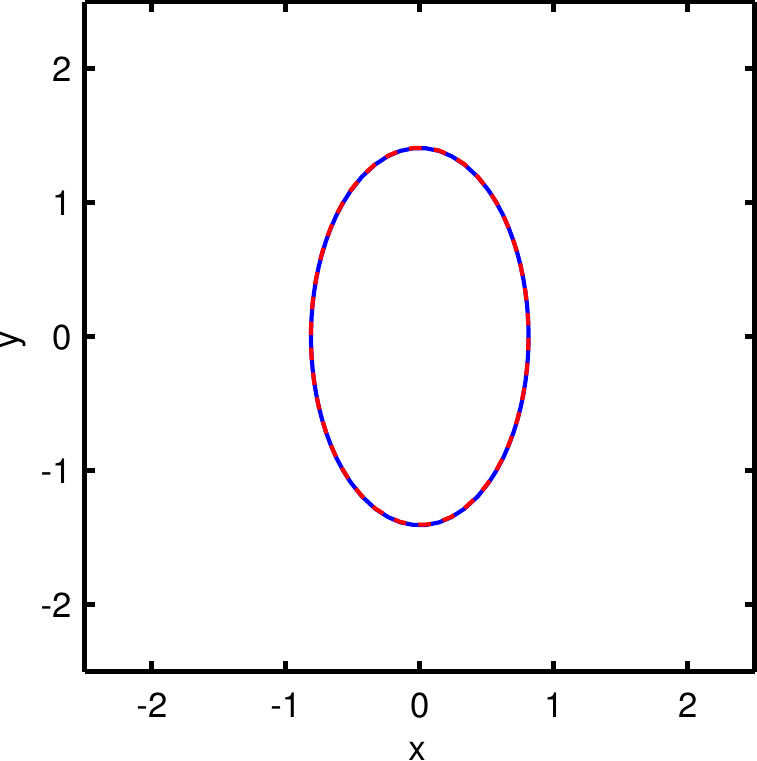}
			\label{fig:XYTBefield00}
		}             
		\subfigure[$t/t_{\dot{\gamma}}=2.5$]{
			  \includegraphics[width=0.15\textwidth]{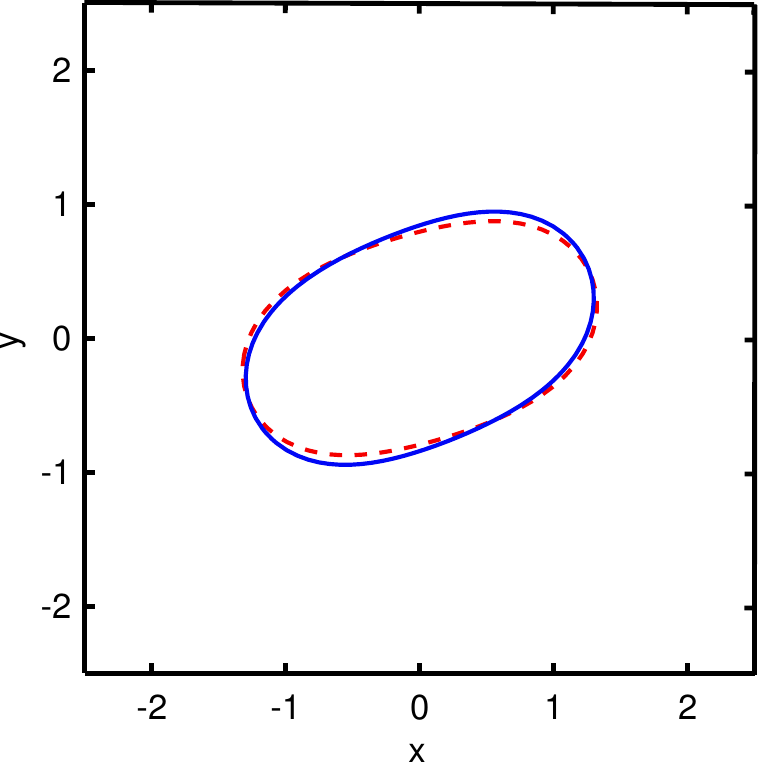}
			\label{fig:XYTBefield025}
		}
		\subfigure[$t/t_{\dot{\gamma}}=3.5$]{
			  \includegraphics[width=0.15\textwidth]{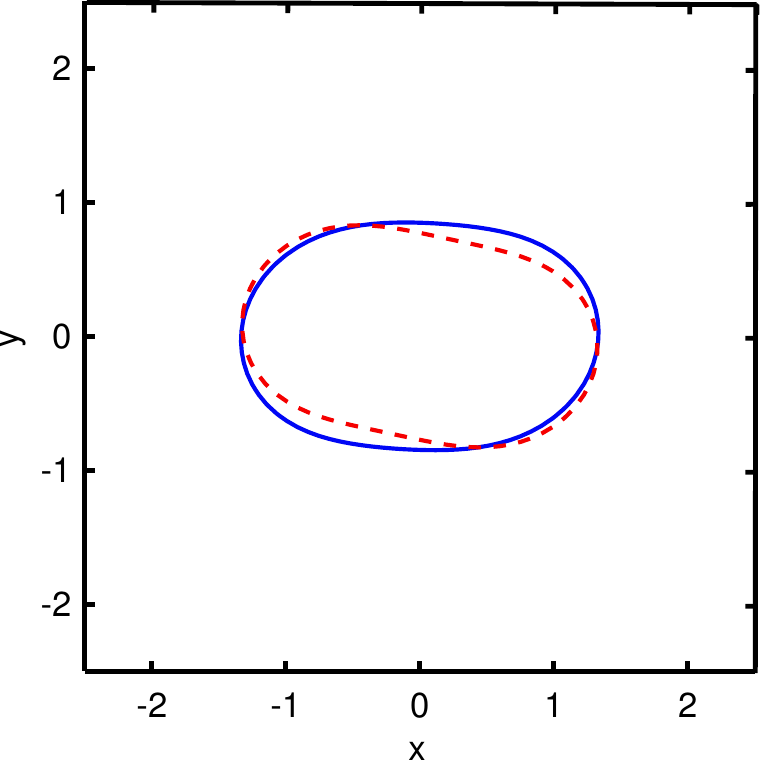}
			\label{fig:XYTBefield050}
		}
		\subfigure[$t/t_{\dot{\gamma}}=6.5$]{
			  \includegraphics[width=0.15\textwidth]{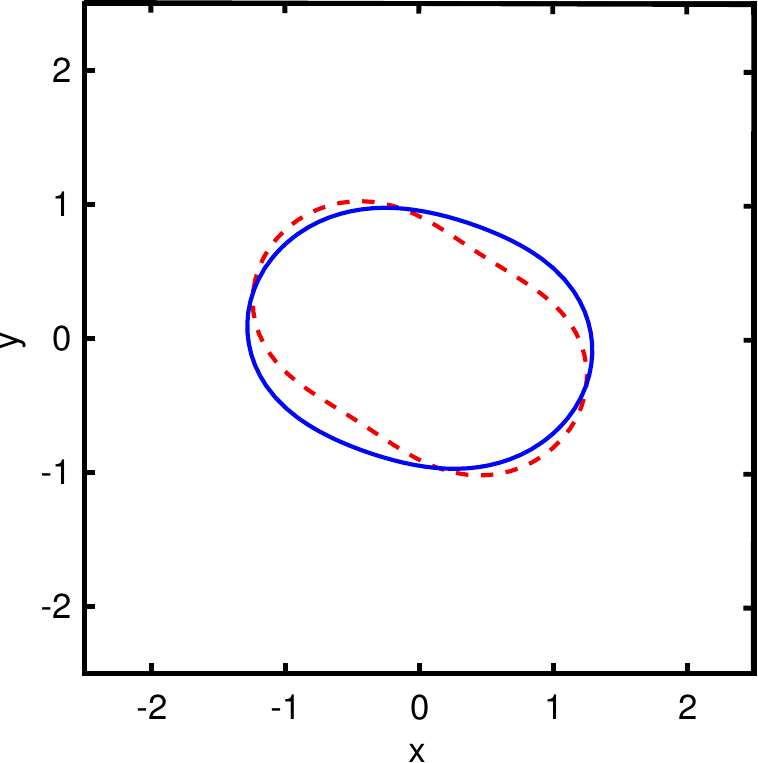}
			\label{fig:XYTBefield065}
		}
		\subfigure[$t/t_{\dot{\gamma}}=7.5$]{
			  \includegraphics[width=0.15\textwidth]{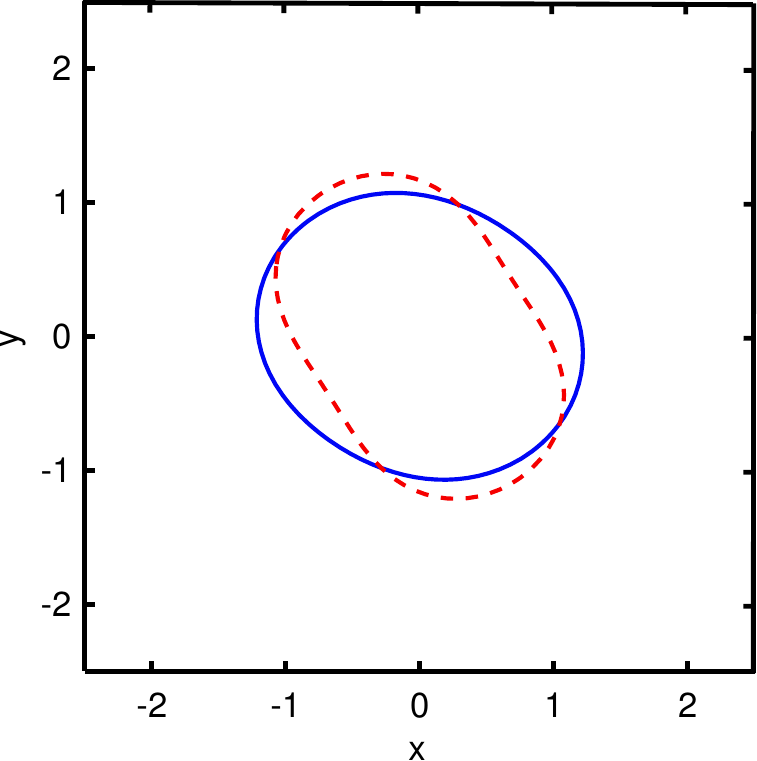}
			\label{fig:XYTBefield075}
		}\\
		\subfigure[$t/t_{\dot{\gamma}}=8.5$]{
			  \includegraphics[width=0.15\textwidth]{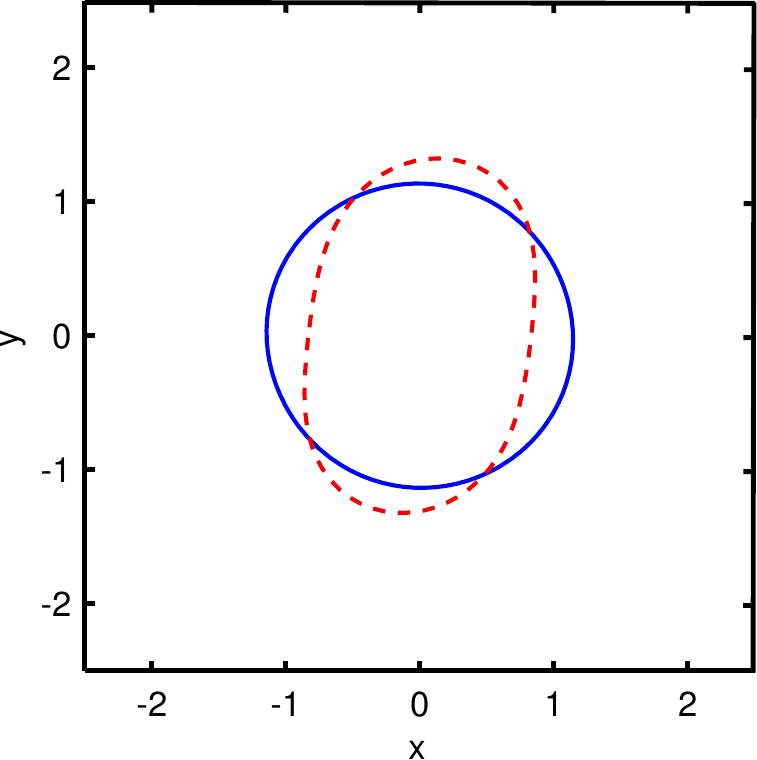}
			\label{fig:XYTBefield08dot5}
		}
		\subfigure[$t/t_{\dot{\gamma}}=10.0$]{
			  \includegraphics[width=0.15\textwidth]{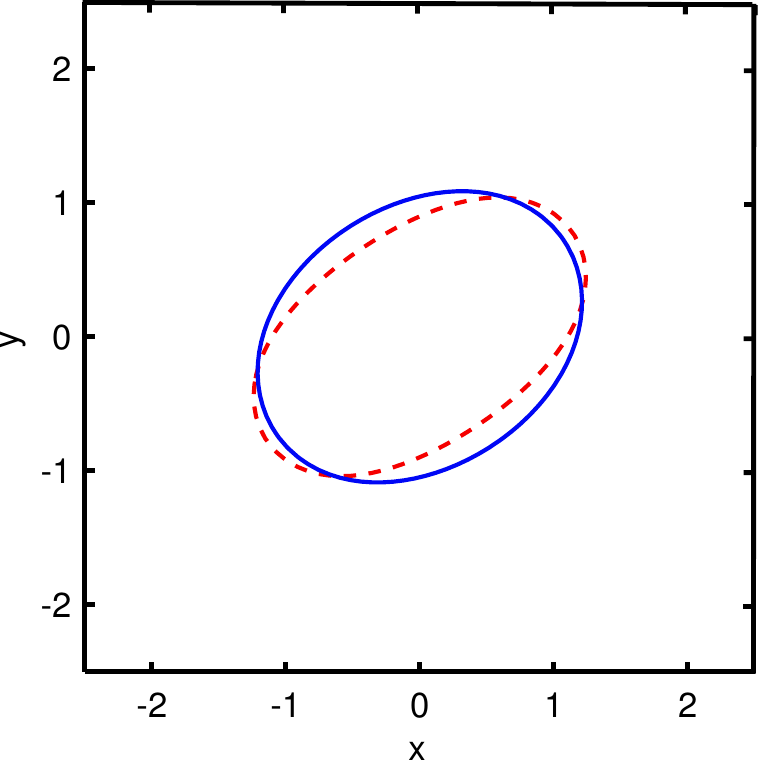}
			\label{fig:XYTBefield010}
		}             
		\subfigure[$t/t_{\dot{\gamma}}=12.5$]{
			  \includegraphics[width=0.15\textwidth]{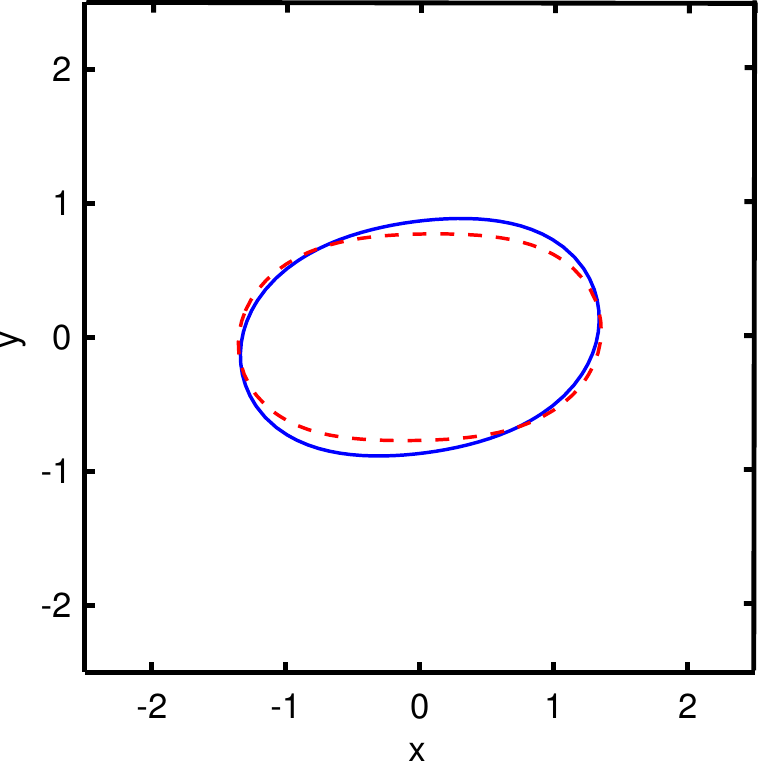}
			\label{fig:XYTBefield012dot5}
		}
		\subfigure[$t/t_{\dot{\gamma}}=15.0$]{
			  \includegraphics[width=0.15\textwidth]{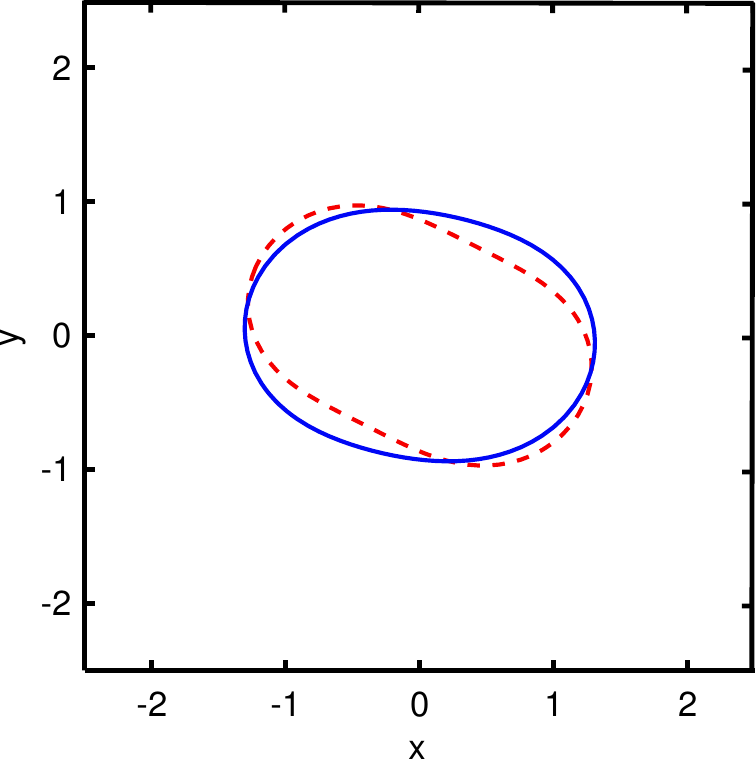}
			\label{fig:XYTBefield015}
		}
		\subfigure[$t/t_{\dot{\gamma}}=17.0$]{
			  \includegraphics[width=0.15\textwidth]{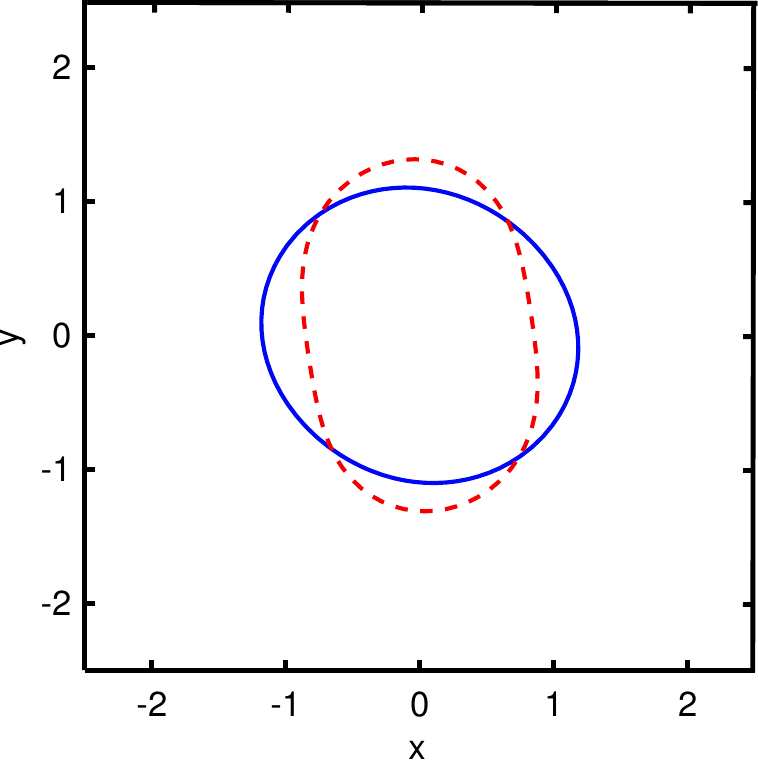}
			\label{fig:XYTBefield017}
		}
		\caption{The x-y cross section of the vesicle with initially charged membrane and $\eta=10$ under combined effect of shear flow and weak DC field 
		with the strength of $Mn=3$ (the solid line; blue in the online version). 
		The dynamics is compared against a tumbling vesicle with the same viscosity ratio and no
		external field, $Mn=0$ (dashed line; red in the online version). The rest of the parameters are $Re=0.001$, $Ca=10$ and $\chi=1$. 
		The presence of the electric field affects the inclination angle and causes the vesicle to reach
		the equilibrium condition at a smaller $\theta$. This behavior is due to the vertical alignment of the electric field in the y direction.
		 The electric field force induces a resistance against any angular movement about the initial vertical position of the vesicle.}    
		\label{fig:XYTB_Efield}
	\end{center}
	
\end{figure}

In the absence of an electric field the vesicle will tumble end-over-end, with the inclination angle undergoing periodic repetition, Fig. \ref{fig:TBefieldAngle}. The application
of a weak electric field, $Mn=3$ results in periodic behavior more akin to trembling. The vesicle does not undergo a rigid-body-like rotation, but instead the 
vesicle poles retract and the vesicle reaches a nearly-spherical shape, see Fig. \ref{fig:TB_Efield} for the three-dimensional representation of the vesicle over time.
A comparison of the vesicles in the $x-y$ plane at the elevated viscosity ratio for the cases of $Mn=0$ and $Mn=3$ is shown in Fig. \ref{fig:XYTB_Efield}. This figure clearly
shows the the tumbling behavior is due to membrane deformation and not and end-over-end rotation.
As the electric field strength is increases to $Mn=15$ the vesicle no longer undergoes a tumbling behavior. Instead the vesicle reaches an equilibrium, tank-treading inclination angle.

\section{Conclusion}
\label{sec:5.0}

In this work a numerical model of the electrohydrodynamics of three-dimensional vesicles has been presented. The work uses a semi-implicit version
of the Jet scheme to implicility track the interface. This is coupled to a full Navier-Stokes projection method with explicit corrections 
to ensure volume and surface area conservation. The electric field contributions are computed through an Immersed Interface Method solver 
which takes into account the jump in the membrane voltage across the interface.

This work was verified by comparing the hydrodynamic results (in the absence of electric fields) to previously published experimental works. The influence
of the time-step, grid-spacing, and domain size on the electrohydrodynamics of vesicles was investigated. The simulation was capable of capturing 
the prolate-oblate-prolate dynamics observed experimentally and predicted by previous analytic and two-dimensional numeric works. The combined 
effects of fluid flow and electric fields were also presented, with weak electric fields dramatically changing the behavior of vesicles in shear flow.

This work is a step towards understanding how to control vesicle dynamics through the use of electric fields and fluid flow. Future work will use the methods
developed here to perform a detailed investigation on how electric fields, fluid flow, and material parameters can be used to influence the dynamics of 
vesicles.

\bibliography{CSF3D}

\end{document}